\documentclass[10pt, showpacs, preprintnumbers, nofootinbib, amsmath, amssymb, aps, prc, twocolumn, groupedaddress,superscriptaddress, floatfix, showkeys]{revtex4-1}

\usepackage{topcapt}
\usepackage[utf8]{inputenc}
\usepackage[dvipsnames]{xcolor}
\usepackage{amsfonts,amsmath,amssymb,bm}
\usepackage{graphicx} 
\usepackage{float}
\usepackage{aas_macros}
\usepackage[pdfpagelabels, pdfencoding=auto, psdextra]{hyperref}
\hypersetup{
 pdfsubject=Paper,
 pdfkeywords={nuclear physics} {Bayesian} {chiral EFT} {scattering},
 unicode = true,
 breaklinks = true,
 colorlinks = true,
 linkcolor = blue,
 citecolor = blue,
 menucolor = blue,
 citecolor = blue,
 urlcolor = blue
}

\usepackage{orcidlink}

\usepackage{longtable}

\newcommand{\numu}{\nu_\mu}
\newcommand{\numubar}{\overline{\nu}_\mu}
\newcommand{\nue}{\nu_e}

\newcommand{\units}[1]{\mathrm{g}/\mathrm{cm}^#1}
\newcommand{\Rnubar}{\overline{R}_\nu}
\newcommand{\Rmubar}{\overline{R}_\mu}

\begin{document}

  \title{Lowest-radiation environments in the Solar System: \\ New opportunities for underground rare-event searches}
  
  \author{Xilin Zhang~\orcidlink{0000-0001-9278-5359}}
  \email{zhangx@frib.msu.edu}
  \affiliation{\href{https://ror.org/03r4g9w46}{Facility for Rare Isotope Beams}, \href{https://ror.org/05hs6h993}{Michigan State University}, East Lansing, MI~48824, USA}

  \newcommand{\uw}{Center for Experimental Nuclear Physics and Astrophysics and Department of Physics, \href{https://ror.org/00cvxb145}{University of Washington}, Seattle, WA, USA}

  \author{Jason Detwiler}
  \email{jasondet@uw.edu}
  \affiliation{\uw}

  \author{Clint~Wiseman}
  \email{wisecg@uw.edu}
  \affiliation{\uw}

\date{\today}

  \begin{abstract}
    We study neutrino, muon, and gamma-ray fluxes in extraterrestrial environments in our Solar System via semi-analytical estimates and Monte Carlo simulations.  In sites with negligible atmosphere, we find a strong reduction in the cosmic-ray-induced neutrino and muon fluxes relative to their intensities on Earth. Neutrinos with energies between 50~MeV and 100~TeV show particularly strong suppression, by as much as 10$^3$, even at shallow depths. The solar neutrino suppression increases as the square of the site's distance from the Sun.  Natural radiation due to nuclear decay is also expected to be lower in many of these locations and may be reduced to effectively negligible levels in liquid water environments.    
    The sites satisfying these characteristics represent an opportunity for greatly extending the physics reach of underground searches in fundamental physics, such as searches for WIMP Dark Matter, neutrinoless double-beta decay, the diffuse supernova neutrinos, and neutrinos from nearby supernova. 
    As a potential near-term target, we propose a measurement of muon and gamma-ray fluxes in an accessible underground lunar site such as the Mare Tranquillitatis Pit to perform a first measurement of the prompt component in cosmic-ray-induced particle production, and to constrain lunar evolution models. Although the focus of this work is on fundamental physics, the finding of this work -- that the existence of an atmosphere in a planet enhances cosmic-ray-induced radiation in its underground -- could be relevant for a broader range of sciences, such as radiobiology deep underground and searches for extraterrestrial life.  
  \end{abstract}

\maketitle

\section{Introduction}

    The past decade has shown a rapid increase in human activities in space from both commercial and government sectors~\cite{2021psad.rept..518H}.
    In the US, for example, the NASA Artemis program is being vigorously pursued in cooperation with private endeavors such as SpaceX and Blue Origin to return humans to the moon and beyond~\cite{smith2020artemis}. 
    Similar programs are being undertaken elsewhere, including in China, India, etc. 
    Rocket launch costs are decreasing rapidly while payload capacities are substantially increasing, potentially leading to a new space era with great scientific opportunities~\cite{Science2024_breakthroughs, jones2018, morganstanley2021investing}. 
    Recently, researchers from fields such as quantum technologies~\cite{belenchia2022quantum}, astronomy~\cite{Serjeant:2020ckp, Jani_2021,Burns:2021pkx, 9438165}, and particle physics~\cite{Zaitsev:2020zui, Beacham:2021lgt,Baum:2024sst,Demidov:2020bff, Solomey:2022ybv,solomey2023neutrino} have started investigating these opportunities.

    In this work, we explore novel opportunities for extending the sensitivity of fundamental-physics rare-event searches by situating them below the surface of various extraterrestrial bodies.
    Such searches include attempts to detect weakly interacting massive particle (WIMP) dark matter by scattering off nuclear targets~\cite{Roszkowski:2017nbc}, to observe the direct creation of matter in the form of neutrinoless double-beta decay~\cite{Agostini:2022zub}, or to search for astrophysical signals such as diffuse supernova neutrinos (DSNs)~\cite{Kresse:2020nto}, or direct neutrino signals from a nearby supernova~\cite{SNEWS:2020tbu}.
    These experiments are being pursued in deep underground laboratories around the world~\cite{Bettini:2011zza} that provide shielding from cosmic rays, primarily muons generated in the Earth's upper atmosphere, that would otherwise swamp those sensitive detectors.

    To extend their reach, these experiments pursue larger detectors with lower backgrounds in each successive generation. 
    In the coming generation of experiments, background due to solar neutrinos is expected to begin to play a significant role. 
    For neutrinoless double beta decay experiments, solar neutrino backgrounds may begin to limit the sensitivity, with a much higher impact for some detection techniques relative to others~\cite{Elliott:2017bui}. 
    In the case of direct dark matter detection, coherent scattering of solar and atmospheric neutrinos, which are born in the same interactions that produce cosmic ray muons, presents an irreducible ``neutrino fog''~\cite{Monroe:2007xp, Strigari:2009bq, Billard:2013qya, Akerib:2022ort, PandaX:2024muv, XENON:2024ijk} common to all experiments, beyond which a background-free experiment is no longer possible even in principle.
    At higher energies relevant for astrophysical neutrino detection, even the deepest underground laboratory experiments suffer from atmospheric neutrino backgrounds. And in all cases, gamma radiation from natural radioactivity is a persistent worry. 

    In this work, we study neutrino,  muon, and gamma fluxes in various underground environments in the Solar System.
    The solar neutrino flux scales as the squared inverse of the Sun-detector distance. 
    Thus, this flux can only be reduced if the detector is moved far from the Sun~\cite{Solomey:2022ybv}. 
    Keeping this in mind, we focus mostly on the cosmic-ray-induced fluxes in extraterrestrial environments. 
    We find that over a broad energy range, the cosmic-ray-induced fluxes can be drastically reduced in these environments. 
    These sites also offer advantages in terms of natural gamma radiation, for which we perform a preliminary investigation.
    
    We examine specifically two archetypical sites: one at shallow depth in a lava tube on the Moon, and another in the form of a human-made ``ice ball'' in space. 
    These sites potentially offer orders-of-magnitude ($10^{3-4}$) reduction of muon and neutrino fluxes relative to what is possible on the Earth, and could enable a new generation of ultra-sensitive experiments capable of detecting new physics that would otherwise remain beyond reach. 
    Our comparative study of the Earth, Moon, and ice ball allows for an analysis of physics possibilities at a number of other, more exotic sites.
    We consider both neutrinos and muons over a large energy range.
  
    It is worth noting that several studies~\cite{Shapiro:1985lbsa.conf..863., Cherry:1985lbsa.conf..863., Petschek:1985lbsa.conf..863., Cherry_10.1063/1.39130, doi:10.1063/1.39125, Cherry_10.1063/1.39130, Wilson_LPSC1998, 2003LPI....34.1392W, 2003ICRC....3.1447W,2006APh....25..368M}, mostly in the 1980s, have schematically estimated cosmic-ray-induced neutrino fluxes on the Moon and explored potential implications for detecting DSNs and high-energy galactic neutrinos there. 
    Recent works~\cite{Gaspert:2023ezv, Solomey:2022ybv} have also discussed the potential relevance of extraterrestrial sites for dark matter detection. In contrast, this work extends Monte Carlo simulations for these underground locations with a comprehensive semi-analytic numerical prediction from un-simulated high energy physics.

\section{Cosmic rays in environments with and without atmosphere} \label{sec:generalphysics}

    \begin{table*}[t]
      \caption{Specific experimental locations considered in this work and their relevant physical parameters. Bold entries are those locations for which Monte Carlo simulations and semi-analytic calculations were performed.}
      \begin{tabular}{lccccc}
        \toprule
        Name & Outer radius [km] & Crust & Crust density [g/cm$^3$] & Surface pressure [atm] & Orbital radius [AU] \\ 
        \hline
        {\bf Earth} & 6400 & rock & 2.8 & 1 & 1 \\ 
        {\bf Moon} & 1700 & rock & \begin{tabular}{@{}c@{}}
            1.6 (regolith)   \\
                2.8 (crust) 
        \end{tabular}  & $3 \times 10^{-15}$ & 1 \\ 
        
        Mars & 3400 & rock & 2.6 & 0.006 & 1.5 \\ 
        Asteroids & 10$^{-3}$--10$^3$ & rock & 1--5 & neg. & 1.5--5 \\ 
        Europa & 1600 & water + ice & 1 & neg. & 5 \\ 
        Rhea & 764 & ice & 1.2 & neg. & 10 \\ 
        Comets & 0.1--30 & ice & 0.3--0.6 & neg. & var. \\ 
        {\bf Ice ball} & 0.1--1 & pure ice & 0.9 & neg. & var. \\ 
        \hline
        \hline
      \end{tabular}
      \label{tab:geometries}
    \end{table*}

\begin{figure}[htbp]
        \includegraphics[width=\columnwidth]{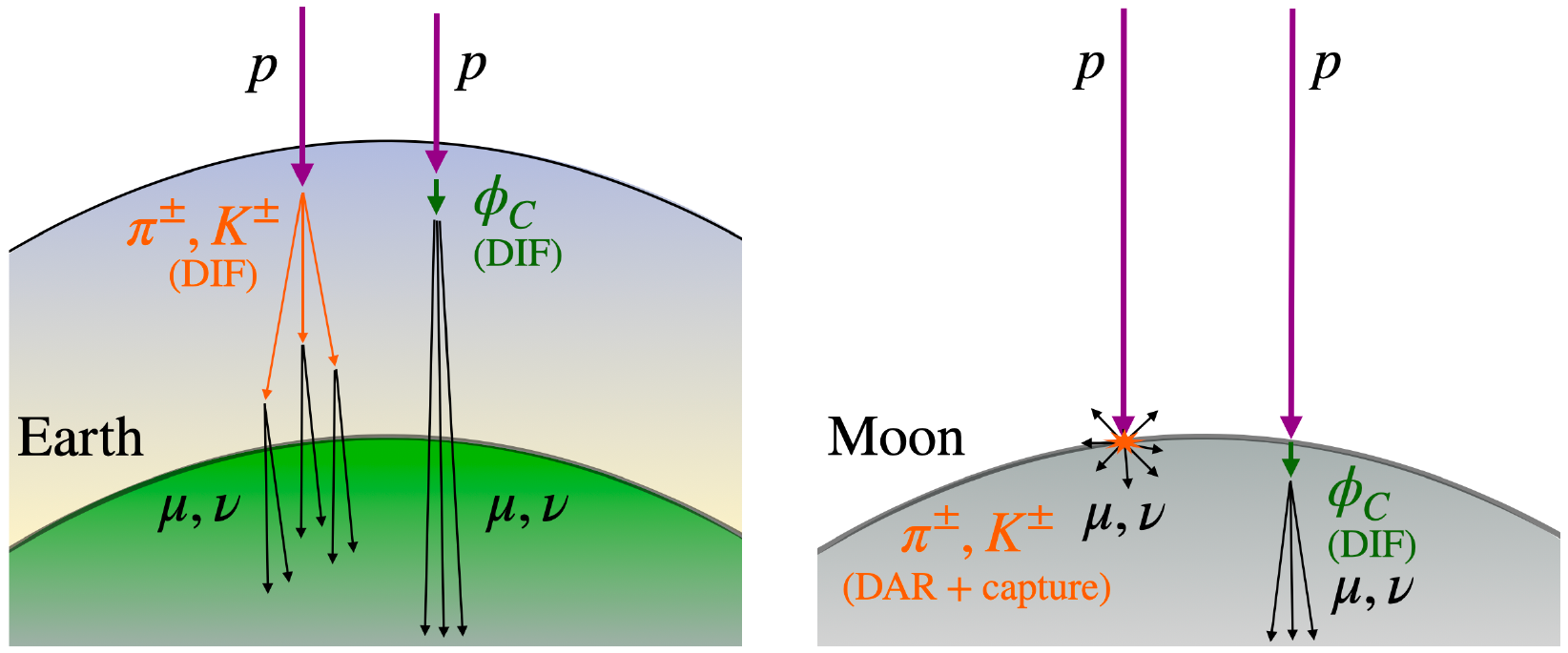}
        \caption{Key neutrino and muon production channels in cosmic ray interactions in the presence or absence of an atmosphere. 
        $\phi_c$ refers to charmed hadrons and similar ultra-short-lived particles that can contribute significantly at very high primary cosmic ray energies.}
        \label{fig:decays_Earth_moon}
    \end{figure}

    Neutrinos and energetic muons created by cosmic rays impinging on matter can penetrate deep underground and generate background in  rare-event experiments. 
    The cosmic-ray induced neutrino flux is accompanied by fluxes from other terrestrial and astrophysical sources, including solar, reactor, and geological (anti)neutrinos and supernovae.
    Most cosmic ray muons and neutrinos are produced far above the Earth's surface, from the decays of short-lived particles generated in hadronic showers, seeded by interactions of primary cosmic rays (mostly protons) with gas molecules in the Earth's upper atmosphere~\cite{Gaisser:2016uoy}. 
    Over most of the energy range that poses background for rare-event searches, the primary production mechanism is from decay-in-flight (DIF) of the charged $\pi$ and $K$ mesons. 
    At the highest energies, an additional ``prompt'' component from the rapid decay of ultra-short-lived particles such as charmed hadrons (collectively labeled as $\phi_c$ here) is expected to dominate. 
    Production of these charmed hadrons by protons is not typically simulated by toolkits like Geant4, so we rely on our semi-analytic estimate of their contribution to the neutrino and muon fluxes.
    Below 50 MeV, an additional mechanism is the decay-at-rest (DAR) of positive mesons and muons that are not captured on nuclei.
    Figure~\ref{fig:decays_Earth_moon} illustrates the primary production channels. 
    
\begin{figure}[htbp]
        \includegraphics[width=0.9\columnwidth]{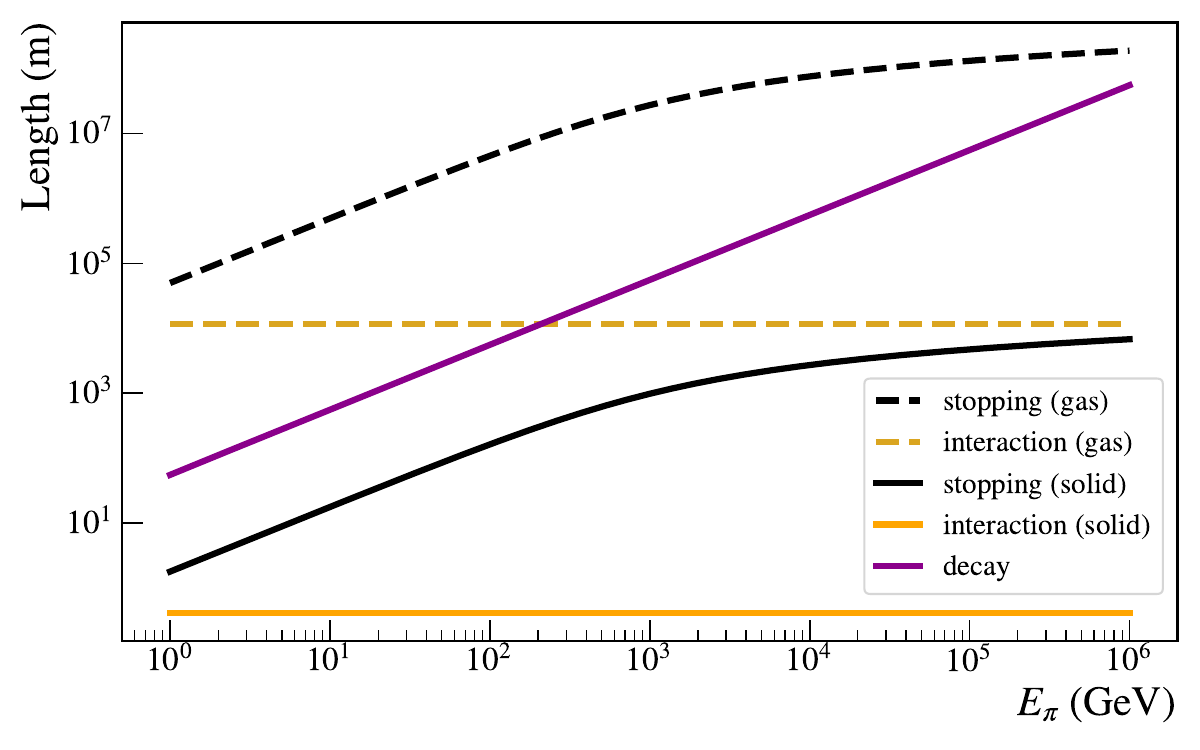}
        \caption{Decay lengths, plotted along with interaction and stopping lengths for energetic pions in gaseous (dotted lines) and solid (solid lines) media, corresponding to densities of $10^{-3}$ and $2.8\ \units{3}$, respectively. Charged kaons show a similar behavior.
        Computed based on Ref.~\cite{Gaisser:2016uoy}; also see Appendix~\ref{app:semiana}.}
        \label{fig:pion_escales}
    \end{figure}

    The presence of a significant contribution to the muon and neutrino fluxes from the meson DIF channel in the atmosphere arises from the fact that in a diffuse gas there is a wide energy range over which these particles' decay lengths are much shorter than the interaction length or the stopping distance, as illustrated in Fig.~\ref{fig:pion_escales}.  
    In environments with no atmosphere, one thus expects qualitatively different secondary cosmic ray production. 
    For primary cosmic rays impinging directly on a solid object with density on the order 1~$\units{3}$, the mesons produced in the hadronic showers have interaction lengths and stopping distances much shorter than the decay length. 
    The DIF muons and neutrinos are thus strongly suppressed in these environments. 
    At low energy, a strong component from the DAR of the positive mesons remains.
    Muons and neutrinos generated by the prompt DIF of ultra-short-lived particles ($\phi_c$) are not expected to be suppressed in environments without atmosphere since the decay lengths of these particles are much smaller than their interaction lengths or stopping distances even in a solid. 
    As a result, the underground fluxes of the highest energy muons and neutrinos are expected to be largely independent of the atmospheric conditions above ground.
    
    The physics of muon and neutrino production in gases versus solids is well established. 
    For example, the ability of dense materials to stop pions is the principle behind the pulsed DAR neutrino source that enabled the first detection of coherent elastic neutrino-nucleus scattering by COHERENT~\cite{COHERENT:2017ipa}. 
    The MiniBooNE experiment used these same DIF and DAR channels to search for exotic Dark Matter~\cite{MiniBooNEDM:2018cxm}. 
    A recent paper~\cite{Demidov:2020bff} also explored how the absence of atmosphere causes the moon to ``shine'' in DAR neutrinos.
    Since energetic $\pi/K$-DIF muons and neutrinos are responsible for much of the cosmogenic background for underground rare event searches, their suppression in environments without atmosphere indicates new opportunities for discovery in extraterrestrial environments. 
    
    In the following sections, we present detailed calculations and Monte Carlo simulations of how these muon and neutrino fluxes differ in environments with differing amounts of atmosphere. 
    We consider three primary simulated environments; Earth, Moon, and the ice ball.
    We also qualitatively discuss secondary environments.
    Key properties of these locations are summarized in Table~\ref{tab:geometries}. 
    We note that the fluxes estimated in this work ignore the suppression and distortion of the cosmic ray flux at low energies due to solar modulation and geomagnetic field~\cite{Gaisser:2016uoy}. 
    As a result, extraterrestrial fluxes in the inner solar system (for example, as calculated in Ref.~\cite{Gaspert:2023ezv}) may be a few times smaller than our estimates.
    
    Given existing efforts aimed at exploring the Moon and Mars, we first consider these locations.
    These bodies are composed mostly of silicates, like the Earth. 
    Five meters of rock amounts to ten nuclear inelastic scattering lengths, so the primary proton flux is reduced by a factor of $\sim\!5 \times 10^{-5}$ over this distance in silicates. 
    This is therefore roughly the minimum amount of shielding needed to absorb all hadronic activity. 
    Ancient lava tubes on the Moon and Mars potentially provide naturally occurring caverns with significant overburden~\cite{SAURO2020103288}.
    In Refs.~\cite{LunaLavaTube:10.2307/24073089} and \cite{LunaLavaTube:GRL}, teams claimed to find lava tubes on the Moon with overburdens of 45 -- 90~m and $>$60~m, respectively. 
    We note with excitement that during the preparation of this manuscript, a report appeared indicating that the Mare Tranquillitatis Pit appears to have horizontal conduits tens of meters long with roughly twice this overburden~\cite{2024NatAs...8.1119C}.  
    Assuming these are not extreme cases, we take the natural length scale of lunar lava tube overburdens to be $\sim$100~m.
    
    Asteroids represent the next-closest target for deployment. 
    Like the moon and Mars, asteroids provide opportunities for scientific endeavors to add value to public- and private-sector interest in mining asteroids for natural resources~\cite{OLeary1977,OSIRISREx2017,Hayabusa22019}. 
    They are located in a range of a few AU from the Sun, providing roughly an order-of-magnitude reduction in the solar neutrino flux. 
    Asteroid compositions vary from carbon-rich to rocky to metallic, with densities of 1.4, 2.7, and 5.3 $\units{3}$, respectively.  
    
    We next consider human-made extraterrestrial low-radiation environments, focusing on ``ice balls''\footnote{``Spaceballs" is perhaps a universal name for variations of this geometry with different materials.} of varying radius. 
    Our concept is motivated by conjectures that water could be extracted from comets in large quantities and used for rocket propulsion~\cite{Sercel8396702, werner2019propulsion}. 
    Combined with purification processes such as filtration and distillation, the extraction of water followed by its controlled solidification could provide a very large and highly radio-pure shield with controllable opacity. 
    We envision deployment beyond the frost line, perhaps in orbit around one of the outer planets, although encapsulation with a highly reflective material could provide stability at lower orbital radii. 
    The ice volume itself could be instrumented as a large neutrino telescope with a design similar to IceCube~\cite{IceCube:2016zyt}, in addition to housing a suite of rare-event searches at its core.
    
    We then consider more distant solar system objects, with a focus on the gas giant moons Europa and Rhea.
    They lie well beyond the frost line near 2.5~AU~\cite{SSWater2008}, providing a strong compositional component of water and ice as well as negligible atmospheric gas densities. 
    Their host gas giants, while large, produce a negligible atmospheric neutrino flux on their moons due to their large orbital radii.
    The surface of Europa, a satellite of Jupiter, is thought to be comprised of 100~km deep ocean covered with a thin crust of $\sim$10~km ice~\cite{EuropaCrust2000}, offering the possibility of achieving significant overburden with limited need for energy-intensive drilling. 
    Rhea, one of Saturn's moons, is likely comprised of mostly ice~\cite{RheaComposition2007}, offering a naturally-occurring, very large version of the ice ball concept. 
    
    Finally, comets, like asteroids, are targets for the mining of resources, in particular volatiles~\cite{SERCEL2018477, Kuck1995}.
    Their often highly elliptical orbits present an opportunity to mount an experiment during a near-Earth approach and let the comet carry the apparatus far from the sun, where solar neutrino fluxes are low. 
    Comet nuclei have a crystalline ice crust and a colder and more porous interior~\cite{CometComposition205}.
    They are surrounded by a gaseous layer (the coma) which becomes larger as the comet approaches the Sun. 
    This layer can be large, but remains diffuse enough that its contribution can be neglected.

\section{Semi-analytic estimates for energetic particle production}

    The formalism for estimating secondary cosmic ray particle fluxes in gases and solids is well established. 
    In particular, the gaseous case is discussed in detail in Ref.~\cite{Gaisser:2016uoy}. 
    These energetic particles, including the secondary mesons and the leptons to which the mesons decay, move effectively collinearly. 
    The changes in their fluxes $N_i$ (particles per unit area, per unit solid angle, per unit energy) along their shared trajectory can thus be described using the following set of coupled equations:
    \begin{align}
      \frac{dN_i(E_i, X)}{d X} = & -\frac{N_i(E_i,X)}{\lambda_i}-\frac{N_i(E_i,X)}{d_i} \notag \\ 
      & + \sum_{j } \int_E^\infty \, dE_j \frac{F_{ji}(E_i, E_j)}{E_i} \frac{N_j(E_j,X)}{\lambda_j} \,.\label{eq:master} 
    \end{align}
    This balances the production of new particles from all available channels (source terms), with their removal via decay or absorption (sink terms) as they traverse matter.
    Here, $X$ is the column depth (distance multiplied by the target mass density), $i$ and $j$ refer to particle types, $\lambda_{i}$ is particle $i$'s absorption length, and $d_{i}$ is its decay length. 
    The term $F_{ji}$ represents the production distribution of species $i$ from parent $j$.

    In our semi-analytic calculations, we numerically solve these equation systems in various environments. A complete discussion of the formalism is given in Appendix~\ref{app:semiana}, including a list of the main symbols and numerical values used (Table~\ref{tab:params}). We consider nucleons, charged pions, and kaons to estimate the muon and neutrino fluxes.
    The neutral mesons mostly decay into photons that are absorbed in the targets and are thus neglected in the calculation. Leptons and their anti-particles are not differentiated unless specified otherwise.  We compute only $\nu_\mu$ flavor of neutrinos since the $\nu_e$ flux is much smaller than the $\nu_\mu$ at high energies in both gases~\cite{Gaisser:2016uoy} and solids.  For the same reasons, the neutral kaons, in particular $K_L$, are neglected here, although they are important for estimating high energy $\nu_e$ fluxes (but not for $\mu$ and $\numu$). 

    \begin{figure*}
        \includegraphics[width=0.9\textwidth]{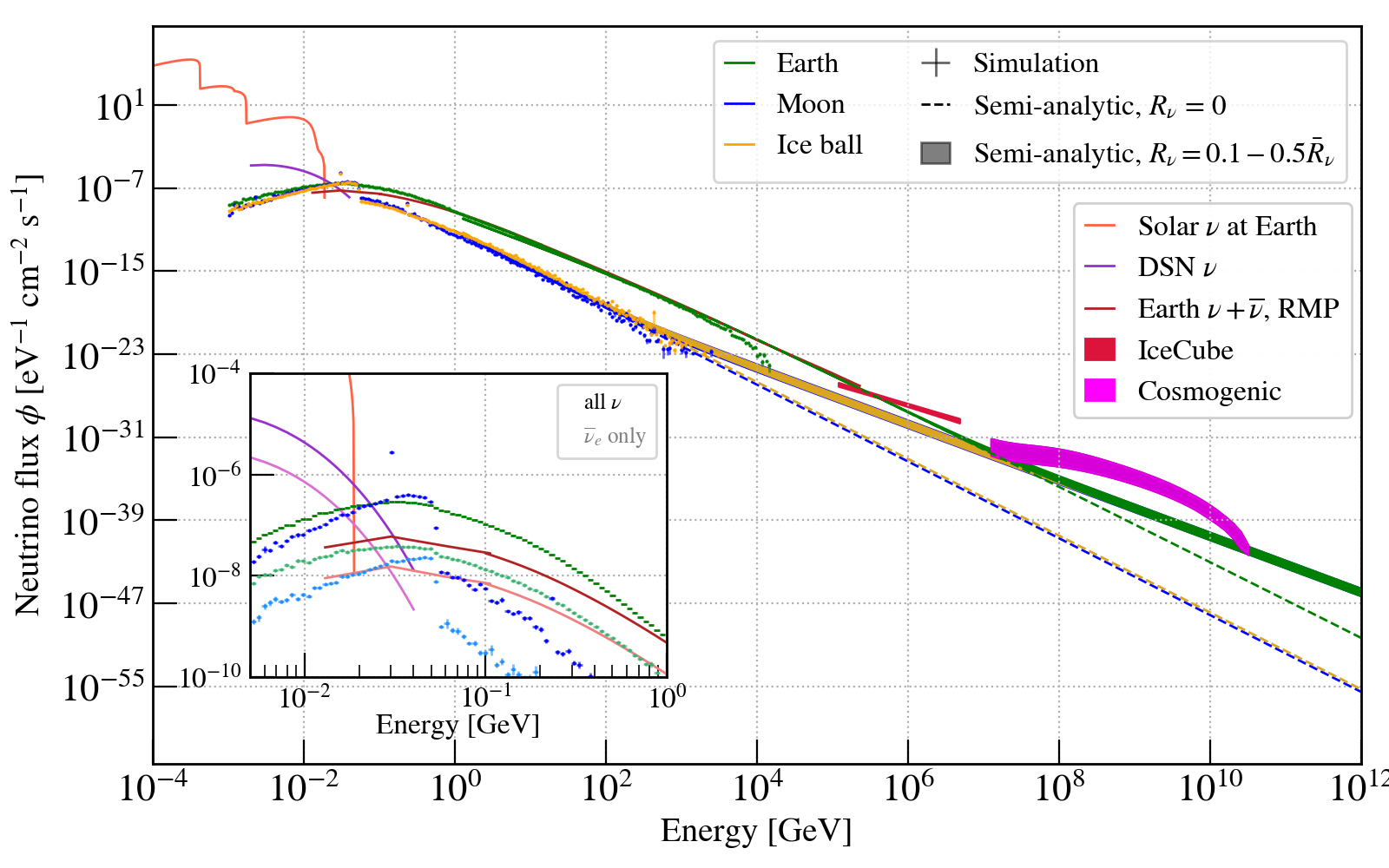}
        \caption{Comparison of our semi-analytic (lines) and Monte-Carlo (crosses, dots in middle) estimations of the neutrino flux at the Earth (1~km depth), the Moon (100~m depth), and the center of a 1~km-radius ice ball (100~m depth), in comparison with flux evaluations in the literature~\cite{Vitagliano:2019yzm,IceCube:2017zho,Moller:2018isk}.
        The ${R}_\nu$ band for the moon lies directly underneath that of the ice ball. 
        \textit{Inset:} Zoom into the 5~MeV -- 1~GeV region for the Earth and the Moon (the ice ball fluxes are similar to those on the Moon), additionally showing the contributions from just the $\overline{\nu}_e$ flavor (drawn in lighter shades), including oscillation.
        }
        \label{fig:nuflux_wide}
    \end{figure*}
    
    Our semi-analytic calculation focuses on the neutrino energy range above 1~GeV. 
    At lower energies, processes like particle stopping, which are not included in the calculations, begin to play a major role. 
    Moreover, solar modulations and geomagnetic fields filter out the cosmic rays with energy below $\sim$10~GeV and thus reduce the secondary lepton fluxes around $\sim$1~GeV~\cite{Gaisser:2016uoy}. 
    We also assume the neutrino mean free path to be infinitely long in matter, which breaks down when the neutrino energy is high enough. For example, the mean free path length inside Earth could be reduced to Earth's diameter when the energy is on the order of $10^5$ GeV (see Fig.~8.5 in Ref.~\cite{Gaisser:2016uoy}). 
    Such energy thresholds are even higher for the Moon and other objects with smaller sizes. 
    In these situations, our results could overestimate the direction-integrated neutrino flux by a factor of two or less due to counting the neutrinos that get absorbed before they could emerge, if possible at all,  from deep underground.   
    
    In addition to the production channels described above, there is also the prompt contribution to the lepton fluxes due to the decays of ultra-short-lived particles such as charmed hadrons.   As briefly mentioned in Sec.~\ref{sec:generalphysics}, their contributions to the lepton fluxes are similar in gases and solids. 
    They have the same energy dependence as the low-energy flux in the atmosphere, which scales with lepton energy as $E^{-(\gamma +1)}$.   Here, $\gamma$ is related to the primary proton flux (c.f.~Appendix~\ref{eq:primary_flux_scaling}), whose energy dependence, to a good approximation, is also $E^{-(\gamma +1)}$.  Therefore, these prompt-component fluxes can be parameterized by a multiplicative rescaling factor ($R_\nu$ or $R_\mu$) applied to the corresponding low-energy flux in the atmosphere~\cite{Gaisser:2016uoy}.  In contrast, the high energy lepton fluxes due to pions and kaons in both gases and solids ($\propto E^{-(\gamma +2)}$) are softer, which is also why the prompt component eventually dominates at high enough energies.  More details about the prompt component are provided towards the end of Appendix~\ref{app:analy_estimate_leptons}.
      
    According to Ref.~\cite{Gaisser:2016uoy}, $R_\nu $ is on the order of $8 \times 10^{-4} \equiv \Rnubar$, while $R_\mu$ should be less than $2 \times 10^{-3} \equiv \Rmubar$. 
    In Fig.~\ref{fig:Numu_Flux_prompt_component_against_pQCD} in Appendix~\ref{app:analy_estimate_leptons}, we compare our prompt neutrino flux results with a recent evaluation based on perturbative quantum chromodynamics (QCD)~\cite{Bhattacharya:2016jce} that satisfies the IceCube upper limit~\cite{Aartsen:2016xlq}. 
    Our results with $R_\nu$ in the range 0.1 -- 0.5 $\Rnubar$ show decent agreement with the pQCD calculations for $E_\nu$ below $10^5$~GeV and above $10^5$~GeV, respectively.

    The estimated neutrino fluxes out to $10^{12}$~GeV are given in Fig.~\ref{fig:nuflux_wide}, where we compare our Geant4 simulations for the Earth, Moon, and ice ball geometries at representative depths, against our semi-analytic calculations and various results from the literature.
    As a benchmark, there is a good agreement between our calculation for the Earth and Ref.~\cite{Vitagliano:2019yzm}. 
    
    In addition to the $R_\nu = 0$ curves, we present the neutrino fluxes with $R_\nu$ varying between 0.1 $\Rnubar$ and 0.5 $\Rnubar$ as a colored band of allowed values. As already mentioned, this $R_\nu$ range is consistent with existing constraints. Due to the prompt component, neutrino fluxes in environments with and without atmosphere converge for energies above $10^7$~GeV or so.

    \begin{figure*}[t]
        \includegraphics[width=0.9\textwidth]{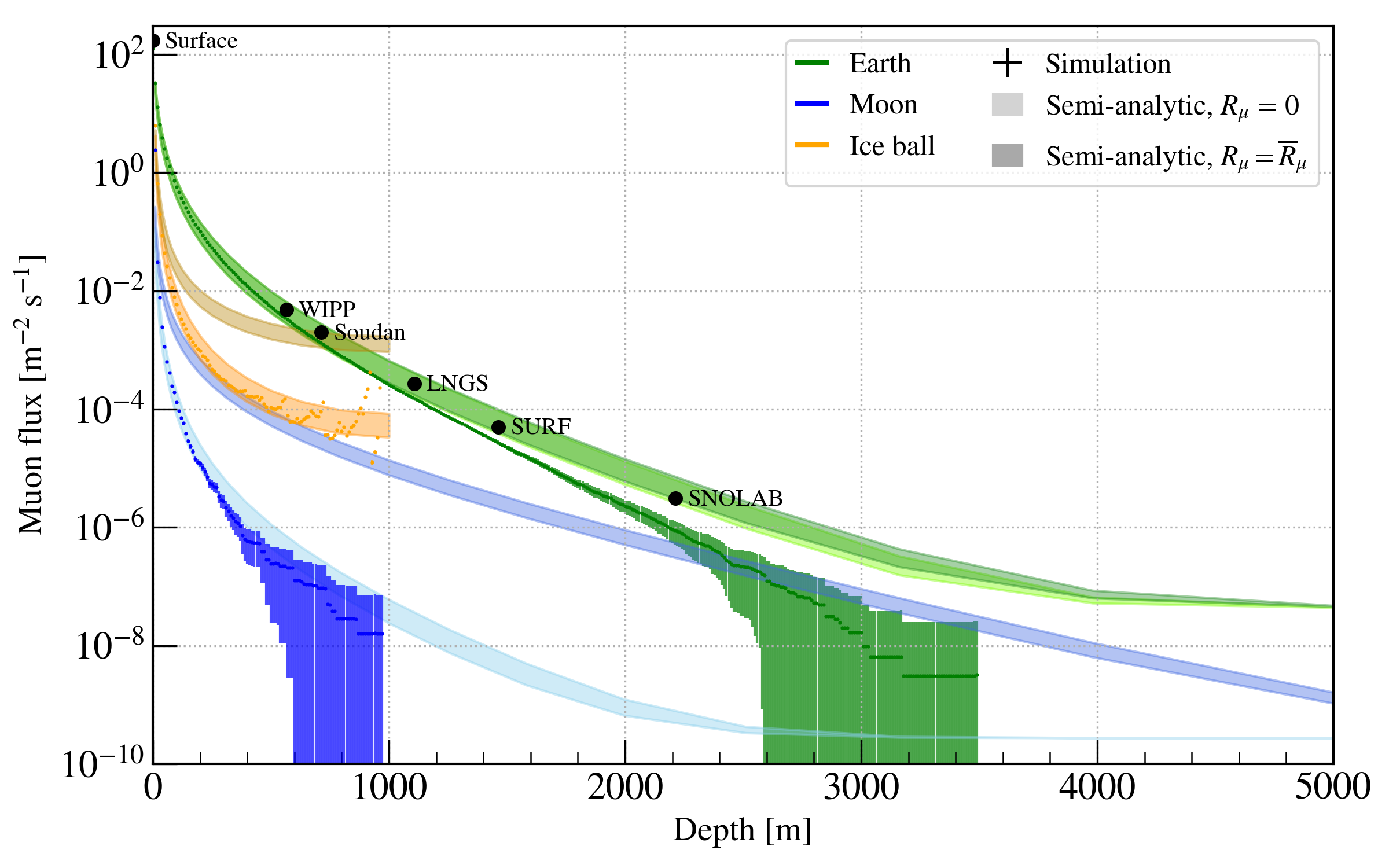}
        \caption{Total muon flux vs.~depth for three geometries, for simulation (crosses, with dots in middle) and for their corresponding semi-analytic predictions with $R_\mu$ = 0 and $\overline{R}_\mu$. The shaded bands correspond to the range $\alpha$ = 2 -- 2.8~MeV~g$^{-1}$~cm$^2$. For comparison, we plot measured muon fluxes at various depths (on Earth)~\cite{ianni2020}.
        }
        \label{fig:muflux_vs_depth}
    \end{figure*}

    We next consider muon fluxes in these locations. 
    Muon energy loss when traversing matter must be taken into account, which can be implemented  by enforcing the $X$ (column depth) dependence of its energy. To model that dependence, we use a standard approximation for the stopping power of penetrating particles~\cite{Gaisser:2016uoy, pdg_2022},
    \begin{equation}
        \frac{dE}{dX} = \alpha(E) + \frac{E}{\xi}\,. \label{eq:muon_E_loss}
    \end{equation}
    Here, $\alpha$ is the energy loss due to ionization, and the second term models radiative processes that rapidly slow particles at very high energies.  More details about muon's energy loss can be found in Sec.~\ref{subsec:muonEloss}. 
    
    Figure~\ref{fig:muflux_vs_depth} shows the energy-integrated muon flux vs depth underground at different environments from our semi-analytical calculations, compared against our Geant4 simulations. 
    The known experimental data for a few underground sites on Earth~\cite{ianni2020} are also plotted as benchmarks.   
    We show the range of the semi-analytic results with $R_\mu$ between $0$ and its upper limit $\Rmubar$. For a given $R_\mu$, we change the $\alpha$ parameter in Eq.~\eqref{eq:muon_E_loss} from 2 to 2.8 -- a range associated with varying muon energy between sub-GeV to 10 TeV -- to get a sense of uncertainty due to $\alpha$'s energy dependence.  
    The Earth estimates are consistent with measurements at various underground laboratories. Although the prompt component has little impact on Earth's muon flux, it is significant in other two environments.  
    For example, within the Moon, the prompt component starts dominating at 100 meters below the surface. 
    Also note that, as shown in Fig.~\ref{fig:muflux_vs_depth}, when the depth is large enough on Earth or Moon, the muon flux flattens to a minimum value, corresponding to the depth-independent neutrino-induced muon flux~\cite{Gaisser:2016uoy}.

\section{Simulation of secondary neutrino and muon production}\label{sec:sims}

    The semi-analytic calculations described in the previous section are accurate for modeling high-energy cosmic ray fluxes where DIF dominates. 
    At low energies where DAR dominates, various physics effects, including particles being scattered off into different directions, mesons' energy loss, and their absorption by surrounding nuclei with nontrivial energy dependence, are highly non-negligible but either not included or not correctly represented in Eq.~\eqref{eq:master}. 
    To account for these, and to compare against established results, we performed Monte Carlo simulations of primary cosmic rays impinging on celestial bodies using Geant4~\cite{GEANT4:2002zbu,Allison:2006ve,Allison:2016lfl}.
    We use Geant4 v4.10.08 with the NuBeam physics list for accurate neutrino production~\cite{abgrall2019nubeam, geant4phys2020}.
    NuBeam is based on the ``FTFP\_BERT'' physics list, which models hadronic interactions below a few GeV with the Bertini intranuclear cascade model and higher-energy interactions using the Fritiof parton model, supplemented with a compound model to de-excite the final-state nucleus. 
    NuBeam uses slightly modified energy ranges over which the Bertini or Fritiof models are invoked, and for high-energy protons above 100 GeV, it switches to a quark-gluon string model. 
    We emphasize that this physics list is not expected to accurately model (if at all) the generation of very short-lived particles (represented by $\phi_c$ in Fig.~\ref{fig:decays_Earth_moon}) that generate the highest-energy muons and neutrinos in the underground flux. 
    Thus, while we expect the simulations to give a more accurate picture at low energies, we expect them to fall below the semi-analytic calculations at high energies.
    
    Planetary geometries were implemented using the Geometry Description Markup Language (GDML)~\cite{Chytracek:2006be}, consisting of concentric spherical geometric shells with appropriately chosen composition and density. 
    The Earth, the Moon, and ice balls with radii 100~m, 300~m, and 1~km were simulated.
    The ice ball geometry was modeled as a simple sphere of solid water ice of density 0.9~g/cm$^3$. 

    The interior of the Earth was modeled following the Preliminary Reference Earth Model~\cite{Dziewonski:1981xy}, except that the average density of the upper crust was changed from 2.6~g/cm$^3$ to 2.83~g/cm$^3$ to account for the fact that physics experiments are performed within the continental crust. 
    The atmosphere was modeled following the U.S.~Standard Atmosphere (1976)~\cite{USStandardAtmosphere1976}.

    For the lunar interior, we used Ref.~\cite{LunarInterior2006}. 
    The crust is mostly plagioclase, modeled here as an even mixture of CaAl$_2$Si$_2$O$_8$ and NaAlSi$_3$O$_8$, with a $\sim$30\% admixture of Al$_2$O$_3$. We include a 7.5-m-thick regolith on the surface of the moon composed of crust material at nearly half its density~\cite{Meyer:2003}.
    The upper mantle comprises orthopyroxene, modeled here as an even mixture of MgSiO$_3$ and FeSiO$_3$. 
    The olivine lower mantle is modeled as an even mixture of Mg$_2$SiO$_4$ and Fe$_2$SiO$_4$. The middle mantle is taken to be an even mixture of orthopyroxene and olivine. 
    Like the Earth, the moon's core is assumed to be 90\% Fe and 10\% Ni.
          
    Primary event generation was performed starting from an incident primary cosmic ray proton. 
    The primary cosmic ray differential flux energy spectrum was taken from the so-called local interstellar spectra (LIS) in which the cosmic ray flux suppression at an energy below 10 GeV due to the solar magnetic field is neglected~\cite{Bisschoff:2015qed}. 
    At higher energy, the flux scales as $E^{-(\gamma+1)}$. 
    Although that spectrum falls steeply with energy, the neutrino and muon production mechanisms increase sharply with energy. 
    To achieve reasonable statistics for the most important processes, primary protons were generated with an energy spectrum flat in $\ln(E)$ between 0.3 GeV and 30 TeV, and results were re-weighted to correspond to the primary differential flux spectrum. 
    The upper limit on the energy range was chosen to avoid excessive computation times and memory overruns in the simulations. 
    The results are essentially unchanged when the lower limit is pushed to lower energies.
    Details on the weighting factors are given in Appendix~\ref{app:sims}.

    \begin{figure}[htbp]
        \includegraphics[width=\columnwidth]{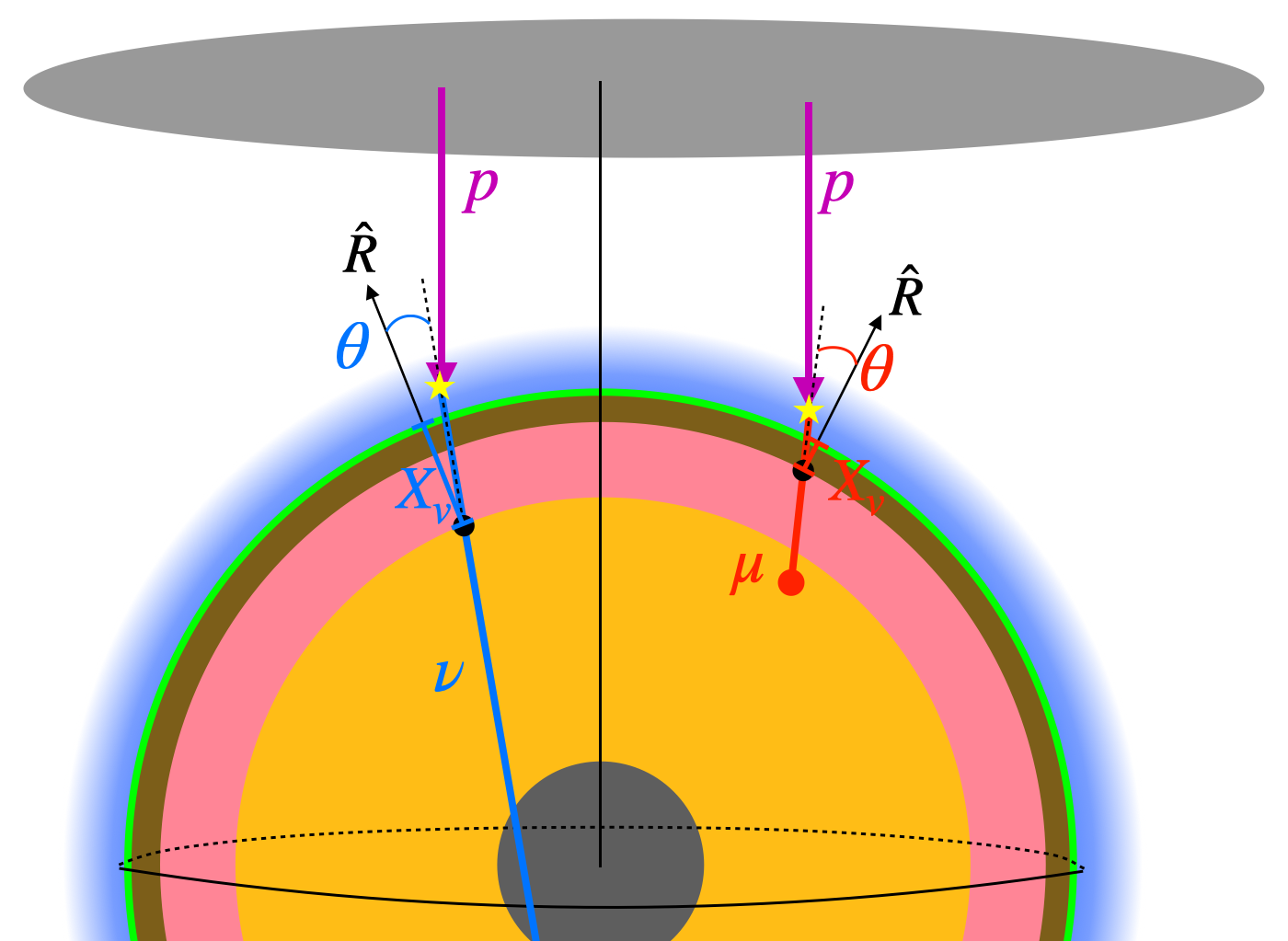}
        \caption{
        Illustration of the methods utilized by our Geant4 simulation, shown here for a simplified version of the Earth geometry.
        Concentric spherical volumes are used for the atmosphere (diffuse blue), surface (green), and interior shells (brown, red, yellow, dark gray).
        Proton primaries (magenta) are generated from a large circular area (light gray) above the outermost shell.
        The tracks of two typical secondary particles are shown, a neutrino (blue) and a muon (red), with the yellow star at the production point.
        For a given shell crossing point (black dot), the particle energy, depth ($d_{\mu,\nu}$) and local zenith angle ($\theta_{\mu,\nu}$) are saved, in order to calculate energy, flux, and for neutrinos, the oscillation amplitude.
        }
        \label{fig:sim-geom}
    \end{figure}
    
    Since we are interested in both muons and neutrinos, it was important to simulate a flux of primary protons impinging isotropically on the body of interest.
    Performing a simulation of cosmic rays impinging isotropically on a planet with a single detector location beneath its surface is highly inefficient since most cosmic rays will not reach the detector location. 
    So, as illustrated in Fig.~\ref{fig:sim-geom}, we instead simulated a uniform flux of primary protons normal to a disk of radius larger than that of the planet (including its atmosphere) and used diagnostic spherical shells at different depths inside the planet to record the fluxes of secondary particles that traverse them. 
    The spherical symmetry of the planetary geometry then allows each traversal, after multiplication by an appropriate weighting factor, to contribute to the statistics for the corresponding depth. 
    This method is inaccurate at low statistics because the traversals in a particular simulated event are correlated. 
    However, such correlations are negligible at the high statistics required to capture the essential low-energy features of the cosmic ray secondaries while matching to the semi-analytic expressions at higher energy.
    
    For each geometry, $10^5$ primaries were generated, and double-different flux of each lepton flavor was computed as a function of depth. 
    At these statistics, tertiary production, e.g.,~of muons generated by atmospheric neutrinos, is negligible. 
    For neutrinos, 3-flavor neutrino oscillation probabilities were applied using the current best-fit parameters in the PDG~\cite{pdg_2022}.
    
    The simulations were benchmarked successfully against measurements and expressions derived for the Earth, found in Ref.~\cite{Gaisser:2016uoy}.  
    We find that the total cosmic ray muon fluxes at the surface of the Earth are reproduced to within 30--40\%, a surprising level of agreement given the approximations used and the exponential showering and attenuation processes involved between the primary proton interactions in the upper atmosphere and the propagation of interacting daughters through the atmosphere. 
    The simulations also reproduce the expected energy and angular distributions well at the surface and depth. 
    
    In Fig.~\ref{fig:nuflux_wide}, we plot the simulation results for neutrinos alongside the semi-analytic computations at different sites. 
    As expected, the Moon simulations reproduce the expected low-energy spectral features due to muon and pion decay that are not included in the semi-analytic calculations, but do not capture the prompt component that is prominent at high energy. 
    The simulations also fall below the semi-analytic calculations above 1~TeV due to the absence in the simulations of contributions from primary cosmic rays with energies above 30~TeV. 
    Nevertheless, between 1~GeV and 1~TeV the two computations show good agreement, so we use them to derive the results and conclusions below. 
    For the Earth's (atmospheric) neutrino flux, the semi-analytic estimates and the result from Ref.~\cite{Vitagliano:2019yzm} all agree with the simulations. 
    However, when the energy is below 1 GeV, the simulation result is above the result from Ref.~\cite{Vitagliano:2019yzm}. 
    This is presumably because our simulations rely on the LIS primary spectrum~\cite{Bisschoff:2015qed}, which does not account for the suppression of the primary cosmic ray flux due to solar modulation and geomagnetic field.  
    
    In Fig.~\ref{fig:muflux_vs_depth}, we compare the underground muon flux versus depth for the Earth, the Moon, and the ice ball. 
    We find the simulation reproduces shallow fluxes well but begin to fall below the semi-analytic calculations at greater depths. 
    The same over-attenuation may be responsible for the suppressed total muon flux at the surface of Earth. 
    This is not unexpected; simulating the exponential attenuation of penetrating cosmic rays is notoriously difficult, even when starting from the measured surface cosmic ray spectrum, and often ad-hoc renormalizations are required (see, for example,~\cite{Kudryavtsev:2008qh}). 
    Considering that we have initiated our simulations from primary protons in the upper atmosphere on Earth or from the surface of the Moon or ice ball, this level of agreement in the absolute flux---within a factor of $\sim$2 for depths of experimental interest---is relatively good.
    
\section{Natural Gamma Radiation}

    To achieve the low backgrounds required for rare event searches, fundamental physics experiments typically must shield against natural radioactivity in the surrounding material. 
    Chief among these is the gamma emissions of the $^{238}$U and $^{232}$Th decay chain species, as well as $^{40}$K. 
    Other contributors include $(\alpha, n)$ reactions from these same decay chains and cosmogenic activation products with long half-lives. 
    We assume here that experimental shielding design will be driven, like it is on Earth, primarily by the need to reduce U/Th/K gamma radiation. 
    The investigation of the other components is left for future work. 
    Natural backgrounds can also take the form of neutrino emissions by U/Th/K species, however we assume here those levels to be negligible; in fact, observing them, along with the gamma ray fluxes themselves, would provide important inputs for planetary formation and composition science.

    With half-lives much longer than the age of the Earth, $^{238}$U, $^{232}$Th, and $^{40}$K were present in the primordial material from which the planets and other solar system objects were born. 
    Chondritic meteorites are thought to be formed by direct accretion of that primordial material. 
    Thus, the concentrations of these species can be taken as expected average levels throughout the solar system. 
    A meta-analysis in~\cite{MCDONOUGH1995223} gives chondritic concentrations of 7.4 ppb U, 29 ppb Th, and 550 ppm (elemental) K (the abundance of $^{40}$K in K is 0.012\%).

    The different evolution histories of various celestial bodies will result in varying primitive concentrations of these species, but perhaps not by many orders of magnitude. 
    A sophisticated estimate of these and other elements in the primitive Earth, incorporating chemical and petrological modeling as well as seismic data, gives a bulk K concentration that is just half of that in chondritic meteorites, while U and Th concentrations are estimated to be around three times higher than their chondritic values~\cite{MCDONOUGH1995223}. 
    However, subsequent processes can lead to significant variation. For example, in the Earth, these elements are excluded from its liquid iron core; convection and geochemical processes result in migration from the mantle to the crust, where concentrations reach 3 ppm U, 10 ppm Th, and 3\% K~\cite{RudnickFountain95}.
    Similar processes occurred on the moon and Mars, but presumably to a lesser extent. 
    Assays of lunar surface rocks retrieved by Apollo missions give concentrations that are lower than values on the Earth by a factor of 7 for U/Th, and by a factor of 50 for K~\cite{TAYLOR1975206}.
    Likewise, investigations of heat-producing elements on Mars based on meteorite data, gamma-ray measurements, and seismic data estimate concentrations of 0.2-0.4 ppm U, 0.9-1.4 ppm Th, and 0.5-0.8\%K in the Martian crust~\cite{2021Sci...373..438K}.

    Concentrations of radioisotopes in water require special consideration. 
    Levels of $^{238}$U in seawater are on the order of 3~ppm, similar to levels in the crust. 
    Th, however, is not soluble in water and settles to the ocean bottom, leaving residual levels of $^{232}$Th on the order of 100 ppt~\cite{CHEN1986241}. 
    And while the upper $^{238}$U chain remains intact, its decay proceeds through other species, including Th, resulting in a strongly broken secular equilibrium. 
    Measurements of $^{226}$Ra, which supports the energetic gamma decay isotope $^{214}$Bi that is a ubiquitous background in rare event searches, indicate concentrations on the order of 0.04 ppq near the ocean surface,
    and higher by a factor of 2-4 near the ocean floor~\cite{Broecker1967}. 
    This is on the order of $10^4$ times less than what would be expected from secular equilibrium with $^{238}$U. 
    The lower $^{238}$U chain ($^{210}$Pb and below) could be out of equilibrium with the mid-chain isotopes. 
    However, the radiation in the lower $^{238}$U chain is typically from alpha or low-energy beta decays, which can be stopped with modest shielding.
    Finally, K has a natural concentration in seawater of 0.04\%~\cite{2024chcp}, corresponding to 48~ppb $^{40}$K. 
    
    Purification of water to ultrapure (18.2 M$\Omega$) levels ($<$0.1 ppt U/Th, 40 ppt K)~\cite{MilliQUPW2019} is possible via processes such as distillation, reverse osmosis, filtration, and degassing, and has been achieved for very large volumes of water for example by the Super-Kamiokande Collaboration~\cite{Super-Kamiokande:2002weg}. 
    With a recirculation rate of 30 t/h, Super-Kamiokande measured residual Rn levels of just 0.2-2 mBq/m$^3$ \cite{Nakano:2019bnr}, indicating that the purification process was achieved without significant injection of Rn or its supporting Ra parent.
    Thus, in principle, radioactive backgrounds can be purified from a liquid water environment to negligible levels.

\section{Results and Discussion}

    In this section we synthesize the results of our semi-analytic and Monte-Carlo investigations for the various extraterrestrial sites of interest. 
    We focus on background reductions relative to terrestrial experiments, with the key results summarized in Table~\ref{tab:summary}. 
    Extrapolation to improvements in sensitivity for specific experimental configurations and physics searches is beyond the scope of this work. 
    We point the interested reader to, for example, the projections in \cite{Gaspert:2023ezv} for lunar WIMP searches, which show improved sensitivity by factors of 2-5 relative to terrestrial experiments, in qualitative agreement with our work. 
    
    Figure~\ref{fig:nuflux_wide} shows that in bodies without atmosphere, even at shallow depths, the cosmic ray neutrino flux between 50~MeV and 100~TeV is $\sim$3 orders of magnitude below that of even the deepest locations on Earth. 
    This result validates our argument made at the outset that the short stopping lengths in a solid body without an atmosphere lead to great reductions in DIF fluxes.
    At low energy, we see in Fig.~\ref{fig:nuflux_wide} and its inset that the simulation over-produces neutrino fluxes on the Earth, as compared to the recommended value from Ref.~\cite{Vitagliano:2019yzm}. 
    As mentioned before, this is likely because our simulations use the LIS flux as the primary cosmic ray flux, which doesn't include the low-energy flux suppression effect due to the solar modulation and Earth's magnetic fields. 
    
    If we specialize to $\overline{\nu}_e$, its $\pi$ DAR creation is highly suppressed thanks to nuclear $\pi^-$ and $\mu^-$ capture processes. 
    However, when including neutrino flavor oscillation, the fluxes between different flavors tend to average out~\footnote{The previous work~\cite{doi:10.1063/1.39125} did not consider the neutrino oscillation effect and thus concluded with strong sensitivity enhancement for DSNs on the Moon.}; the $\bar{\nu}_e$ flux at 100 m depth presented in the plot is only a factor of a few smaller than the flux of Earth's atmospheric $\bar{\nu}_e$. 
    Meanwhile, our simulation does not have the solar-modulation-induced suppression on primary cosmic ray flux. 
    Therefore, in the bodies located in the inner solar system, $\bar{\nu}_e$ could be smaller than our simulation. 
    For example, based on the ratio between our simulation of Earth's atmospheric neutrino flux and the recommended value~\cite{Vitagliano:2019yzm}, we can infer the impact of the solar magnetic field. 
    Including that factor, we expect that there may be some enhanced sensitivity (around a factor 10) to DSN $\bar{\nu}_e$ signals in such environments without atmosphere. 
    However, given the uncertainties in our calculations, this conclusion is only tentative and requires further study.
    
    The three bodies explored in Figure~\ref{fig:muflux_vs_depth} show qualitatively different muon flux dependence on depth. 
    The underground muon flux on Earth, down to depths of several km, is dominated by high-energy muons generated during decay-in-flight of atmospheric hadronic shower secondaries.
    In the two bodies without atmosphere, we confirm the expected rapid suppression of those DIF muons in the first $\sim$100~m of depth. 
    At greater depth, however, if $R_\mu$ is near its maximum value $\overline{R}_\mu$, we expect the prompt component to become dominant quickly.
    In the case of the ice ball, the lower density of ice leads to a less-rapid flux suppression with depth, and the flux plateaus near the center of the object due to geometrical effects. 
    Still, we find that even with a maximal prompt component, one can achieve cosmic ray muon rates equivalent to those of several of the Earth's deepest laboratories at depths of only $\sim$100~m or greater on the moon or at the center of a 1~km ice ball. 
    We further note from Fig.~\ref{fig:muflux_vs_depth} that bodies without atmosphere show a reduced ``muon floor,'' where the muon flux plateaus at the deepest depths due to neutrino-induced muon production. This reduction is a direct consequence of the reduced DIF neutrino fluxes at medium energies visible in Fig~\ref{fig:nuflux_wide}. For bodies without atmosphere, this muon floor is expected to scale as the inverse of the density at the body's surface since a higher surface density results in more rapid stopping of energetic particles. 
    
    \begin{table*}
        \caption{Neutrino, muon, and gamma-ray flux suppression factors at different extraterrestrial sites relative to Earth. Here we assume $R_\nu = 0.1-0.5 \overline{R}_\nu$ and $R_\mu = \overline{R}_\mu$. ``Low-'' and ``mid''-energy $\nu$ refer to neutrinos with $20<E_\nu<50$~MeV and 50~MeV $<E_\nu<100$~TeV, respectively; higher energy neutrinos are dominated by the prompt component that is similar for all environments. Muon suppression is expressed as an equivalent muon flux depth on the Earth, $D_\mu$. }
        \begin{tabular}{lccccccc}
            \toprule
            Name & solar $\nu$ & low-$E$ $\overline{\nu}_e$   & mid-$E$ $\nu$& $D_\mu$ & U & Th & K \\ 
            \hline
            Lunar pit ($\sim$100~m) & 1 & 3-10 & $10^3$ &  1~km & 7 & 7 & 50 \\ 
            Mars lava tube ($\sim$100~m) & 2 & 1  & 1   &  100~m & 10 & 10 & 5 \\ 
            Asteroid (100~m bore) & 2-25 & 3-10 & $10^3$  &  1~km & 100 & 100 & 100 \\ 
            Europa ($>$10~km bore + sub.) & 25 &  3-10 & $10^3$ &  $>$10~km & 10$^4$ & 10$^5$ & 100 \\ 
            Rhea (1~km ice bore) & 100 &  3-10 & $10^3$   &  1~km & 100 & 100 & 100 \\ 
            Comets (1~km ice bore) & var. &  3-10 & $10^3$   & 1~km  & $>$100 & $>$100 & $>$100 \\ 
            Ice ball (1~km radius) & var.  &3-10 & $10^3$  & 700~m & $>$10$^4$ & $>$10$^8$ & 10$^9$ \\ 
            \hline
            \hline
        \end{tabular}
        \label{tab:summary}
    \end{table*}
      
    These considerations along with our earlier discussions can be used to estimate expected reductions in cosmic muon, neutrino, and gamma ray fluxes in all of our sites of interest.  
    We summarize the results in terms of suppression factors in Table~\ref{tab:summary}. 
    For muons, we express the suppression in terms of the equivalent depth on the Earth $D_\mu$ that has a similar muon flux. 
    For neutrinos in addition to solar neutrinos, we discuss cosmic-ray-induced neutrinos in two different energy ranges. 
    At low energies, the fluxes are mostly from DAR and have little depth dependence. 
    At higher energies, the fluxes are generally suppressed by a factor on the order of $10^3$ -- $10^4$, which is the ratio of gas density and solid density. 
    At even higher energies, the prompt component starts dominating in these sites, and eventually, the suppression factor decreases to $\sim$1 at high enough energy. 
    The exact critical energies are not clear because they depend on the not-well-constrained prompt component. 
    We discuss each site of interest in the following.
    
    The moon represents a great opportunity for fundamental physics measurements. 
    Not only is it the closest celestial body with no atmosphere, as mentioned earlier, but it is also confirmed to have existing lava tubes with overburdens on the order of 100~m of rock. 
    An experiment could thus be fielded without significant excavation, a major advantage. 
    Atmospheric neutrino backgrounds would be much lower than those on Earth. 
    At low energy, a few 10s of MeV, the $\bar{\nu}_e$ shows some suppression, opening a window for DSN observation.
    Depending on the strength of the prompt contribution, the muon flux will be equivalent to or even lower than those of the deepest laboratories on Earth. 
    An exploratory mission, for example, at the Mare Tranquillitatis pit~\cite{2024NatAs...8.1119C} to measure or constrain this prompt component would be a valuable and feasible endeavor. 
    This component has been elusive in experiments on Earth but is important to particle astrophysics and high-energy nuclear physics~\cite{Gaisser:2016uoy}.
    Simultaneously measuring gamma-ray fluxes in the environment would also provide an interesting cross-check of lunar surface rock radioassays.
    
    For Mars, we do not expect significant suppression of the neutrino and muon fluxes (see Appendix~\ref{app:semiana} for a detailed discussion). 
    Even though the vertical column depth on the surface of Mars is only $\sim$1/100$^{\rm th}$ that of Earth, its absolute value, $150\,\units{2}$, is similar to what is needed for pion fluxes to develop to their maxima. 
    Meanwhile, the column depth slowly varies with physical height $h$ above the surface according to $X_v = X_0 e^{-h/h_0}$. 
    The typical length scale $h_0$ is similar to that on Earth ($\sim$6.5~km). 
    This indicates that on the order of a few kilometers above the surface of Mars, the pion flux reaches its maximum. 
    Within such a long distance, pions with energy below 100~GeV will decay and produce energetic neutrinos and muons. 
    As a confirmation, a detailed Geant4 simulation of the radiation near Martian surface in Ref.~\cite{Dartnell2007_MartianRad} shows similar surface muon rate (see the Fig.~2 therein) as the surface rate on the Earth (see Fig.~\ref{fig:muflux_vs_depth} at zero depth).
    However, a gamma-ray flux measurement in a Martian lava tube would still provide an interesting input to planetary formation models and potentially help resolve uncertainties in Martian structure analyses~\cite{2021Sci...373..438K}.
    
    Asteroids represent the nearest bodies that offer a significant reduction in solar neutrino fluxes by factors of 2-25 depending on their solar orbital radius. 
    Their density and size show a large range of variation, but muon and low/mid-energy neutrino flux suppression factors between those of the Moon and our ice ball geometry can be expected. 
    For gamma radiation, one anticipates levels similar to that of the primordial Earth, offering roughly two orders of magnitude reduction relative to the Earth's crust. 
    We also note that the low gravity on asteroids makes excavation much less energy-intensive than on Earth.
    A preliminary mission, perhaps piggybacking on a private-sector endeavor, to measure muon and gamma-ray fluxes would have substantial scientific value.
    
    A surprise finding is that Europa potentially provides a significant reduction for all cosmic and gamma ray fluxes relative to the Earth, provided one can bore through its $\sim$10~km solid-ice crust in surface temperatures of 50-100~K. 
    Drilling into ice on scales beyond 1~km has at least already been achieved by the IceCube Collaboration~\cite{2014AnGla..55..105B}. 
    A detector submersed in the liquid water just below the crust would have suppressed cosmic ray neutrino fluxes similar to those on the moon. 
    Muon fluxes would be at Europa's muon floor, roughly twice that of the Moon's and much lower than levels achievable on Earth. 
    The gamma radiation from U and Th would be reduced by many orders, leaving a dominant contribution from $^{40}$K that is still two orders of magnitude lower than that on Earth. 
    Solar neutrinos would be suppressed by a factor of 25. 
    As a distant body containing liquid water, there is a very exciting possibility that Europa harbors life, implying an opportunity for piggyback missions in astrobiology, but accompanied by stringent sterilization requirements so as not to contaminate the pristine Europa ecosystem.
    
    Rhea, like Europa, is distant from the sun and has an icy composition. 
    While Rhea offers an additional factor of four lower solar neutrino background over Europa, it is most likely made of a mixture of rock and ice, and does not appear to harbor a large liquid water volume below its surface. 
    It is thus unlikely that boring can be achieved by simply melting the ice, and the gamma ray fluxes are likely to be similar to those of asteroids.
    
    Comets are thought to be composed mostly of frozen ammonia, methane, carbon dioxide, and water mixed with rock, and thus would provide a composition perhaps similar to Rhea's in terms of boring technique, but likely with improved natural radiation levels. 
    A comet with a highly elliptical orbit could potentially transport an experiment much further from the sun, achieving a much lower solar neutrino flux. 
    A comet-based science endeavor could piggyback on a mining mission. 
    The demonstration of the ability to harvest water from a comet could simultaneously achieve commercial goals while laying the groundwork for a future ice ball endeavor. 
    However, highly elliptical comets are naturally volatile environments, and one would need to deal with instabilities and coma generation while the comet is near the sun.
    
    Should water harvesting prove viable, one can imagine subsequently purifying the water and using it to fill a giant balloon surrounding an experimental apparatus and then slingshotting the apparatus into a highly elliptical orbit. 
    A reflective balloon material would help minimize the impact of solar radiation, and the water would freeze on its own before the balloon reaches the frost line on its way to the edges of the solar system. 
    Such an ice ball would provide all of the advantages of Europa in terms of low cosmic neutrino fluxes, but with potentially greatly reduced solar neutrino and gamma ray fluxes, at the cost of a somewhat higher muon flux. 
    In addition to its scientific value, such an object would also serve as a water reservoir in outer space, a potential asset.

    Finally, although this study is focused on implications for rare event search experiments, its counterintuitive central finding that the presence of an atmosphere enhances cosmic-ray-induced radiation underground, could be important to studies outside of physics, such as the search for extraterrestrial life~\cite{merino2019living} and underground radiobiology~\cite{belli2020response, lampe2017understanding, liu2018history}. 
    For example, radiobiology studies are being performed in various terrestrial underground laboratories to take advantage of their low level of radiation; we therefore expect this work to be potentially relevant there.

\section{Comparison with previous studies} \label{sec:existingworks}
    
    In the 1980s and early 90s, several studies addressed comic-ray-induced neutrinos on the Moon and the potential advantage of hosting neutrino detection experiments there. 
    In the energy region above GeV, mostly via simple estimations, Refs.~\cite{Shapiro:1985lbsa.conf..863., Cherry:1985lbsa.conf..863., Petschek:1985lbsa.conf..863., Cherry_10.1063/1.39130} came to qualitatively similar conclusions as we have about the suppression of the lunar neutrino flux relative to the atmospheric neutrino flux on Earth~\footnote{Ref.~\cite{Petschek:1985lbsa.conf..863.} made an interesting additional observation: the Earth subtends a solid angle of $5 \times 10^{-5}$ out of $4\pi$ on the Moon, so the Earth's atmospheric neutrino flux that reaches the Moon is suppressed by $10^{-5}$ to $10^{-6}$ relative to its magnitude on Earth.}. 
    Later, Ref.~\cite{2006APh....25..368M} performed more detailed semi-analytical calculations of neutrino fluxes in the energy region above 1~GeV. 
    The results are qualitatively similar to ours. 
    
    In the energy region below GeV, Ref.~\cite{doi:10.1063/1.39125, Cherry_10.1063/1.39130} explored the potential advantage of detecting DSNs on the Moon. 
    However, their estimates of cosmic-ray induced neutrino fluxes~\cite{doi:10.1063/1.39125}, including the full and $\bar{\nu}_e$ fluxes are qualitatively different from ours. 
    Their full neutrino flux was assumed to be the same as the flux at 1 GeV; when estimating $\bar{\nu}_e$, the neutrino oscillation effect, which tends to average out the difference between electron and muon flavors, were not included. 
    Consequently, their cosmic-ray-induced $\bar{\nu}_e$ flux on the Moon was underestimated. 
    Several MC simulations were also performed~\cite{Wilson_LPSC1998, 2003LPI....34.1392W,2003ICRC....3.1447W, 2003LPI....34.1392W} for the cosmic-ray induced neutrinos on a cylinder solid target, but those flux results, which are several orders of magnitude smaller than ours, do not directly apply to the Moon. 
    
    Most previous work has focused on the Moon hosting neutrino detection experiments at both low and high energies. 
    The potential relevance to other rare event detection experiments is typically not discussed.
    We note Ref.~\cite{Learned_10.1063/1.39124} mentions dark matter, proton decay, double beta decay, and the prompt component detection on the Moon, but does not provide quantitative estimates of muon fluxes.  

    Recently, Ref.~\cite{Demidov:2020bff} simulated the cosmic ray radiation on the Moon, exclusively focusing on observing the neutrino fluxes from the moon on the Earth. 
    The authors carefully studied the impact of such neutrino background on dark matter detection.
    Their results were used in Ref.~\cite{Gaspert:2023ezv} to infer the neutrino fluxes inside the Moon at a depth of 1 km.  
    However, our results, including both simulation and semi-analytical estimations, imply somewhat greater suppression factors, possible arising from the neglection in this work of solar modulation and geomagnetic fields.
    According to~\cite{Gaspert:2023ezv}, at 0.1 and 1~GeV, the neutrino flux ratios between the Earth and the Moon are factors of 10 and 100 respectively, a factor of $\sim$3 smaller than the ratios in this work. 
    Also, the approach of reducing solar neutrino flux via moving detectors further away from the Sun was discussed in Ref.~\cite{Solomey:2022ybv}; however, the study did not address muon fluxes and cosmic-ray-induced hadronic showers.
    
    In comparison, our work conducts a study of neutrinos and muons in extraterrestrial sites. 
    We report neutrino and muon fluxes at characteristic depths over a large energy range, and provide additional information on the origin of various types of background signals (cosmic-ray induced, natural radiation, and solar neutrinos).
    Additional sites of interest beyond the Moon are discussed.
    We note again that previous work has not considered underground muon fluxes in these locations. 
    So far, this is the only work that implements both semi-analytic calculations and computer simulations, benchmarks them in the overlapping energy ranges, and then extends the energy ranges using the two approaches.

\section{Conclusions}

    We have investigated solar neutrino, cosmic neutrino, cosmic muon, and natural gamma-ray fluxes in various extraterrestrial sites of scientific interest. 
    Our results for suppression factors in these fluxes over those of underground laboratories on Earth are summarized in Table~\ref{tab:summary}. 
    A combination of semi-analytical calculations and Monte Carlo simulations were employed to understand the cosmic-ray-induced neutrino and muon fluxes. 
    We achieved good agreement between these calculations at medium neutrino energy ranges and shallow depths. 
    The Monte Carlo simulations helped us understand the low-energy neutrino and muon fluxes that can be dominated by decay-at-rest signatures in bodies without atmosphere. 
    The semi-analytical approach provided a handle on the prompt component contributions that are missing in the simulation physics. 
    
    Generally speaking, sites without atmosphere show a suppression of cosmic neutrino fluxes on the order of $10^3$ -- $10^4$ in the mid-$E$ range. 
    At the low-$E$ region, the total neutrino flux is not substantially suppressed, but we expect $\bar{\nu}_e$ flux to be suppressed by a factor of $\sim$10 after including neutrino flavor oscillation.  
    The underground muon fluxes in these environments are strongly suppressed as well. 
    For example, 100~m deep in the moon, the muon flux is similar to that at a depth of 1~km in the Earth. 
    At the deepest depths, the muon fluxes approach a neutrino-induced floor. 
    Again, this minimum is drastically reduced at sites without atmosphere, e.g., by a factor of $\sim$1000 if we compare the Moon to Earth. 
    The natural radiation at most extraterrestrial sites also shows promising potential reductions compared to that on the Earth. We found that it can be reduced to essentially negligible levels, particularly in water environments. Distant bodies or bodies in highly elliptical orbits can also provide greatly reduced solar neutrino fluxes simply due to the $1/r^2$ scaling of that flux.
    
    In these environments, significant sensitivity improvements can be achieved for rare-event searches such as Dark Matter direct detection and neutrinoless double-beta decay, for which those fluxes pose problematic backgrounds. 
    We also note that lower natural radioactivity implies greatly reduced gamma ray shielding requirements. 
    Of the sites considered, Jupiter's moon Europa, along with a human-made ice ball in a highly elliptic orbit emerged as the two most potentially advantageous sites for scientific discovery.
    
    Considering the obvious challenges of performing rare detection experiments in such sites, we propose a successive approach. 
    A near-term project could measure the muon and gamma-ray fluxes in Lunar lava tubes by piggybacking on current Moon exploration efforts. 
    Such a mission would discover and constrain the prompt component of cosmic-ray-induced particle production, which is important for particle physics and astrophysics, and could also measure gamma radiation to constrain planetary evolution models. 
    Science missions performed jointly with asteroid or comet mining expeditions could provide similar important measurements for those bodies, while simultaneously demonstrating drilling and water harvesting technologies that could enable future missions.

\section*{Acknowledgments}
    X.Z.~would like to thank Professor John Beacom for pointing out the 1980s studies on lunar neutrino fluxes when X.Z.~was working as a postdoc at the Ohio State University. We thank B.~Zhu for useful feedback on our work. We also thank M.~Stortini for early contributions to the simulations. This material is based upon work supported by the U.S. Department of Energy, Office of Science, Office of Nuclear Physics, under the FRIB Theory Alliance Award No. DE-SC0013617 (for X.Z.), and under Award Number DE-FG02-97ER41020 (for J.D. and C.W.), which also supported computation for the simulations on the CENPA Rocks cluster.

\appendix

\section{Simulation geometry and weighting factors} 
\label{app:sims}

    Our simulations generate a large number of proton primaries for various geometries, including the Earth, Moon, and ice ball. 
    From these primaries, we examine the secondary muons and neutrinos, which reach a laboratory depth of interest. 
    We then evaluate the flux of secondary muons and neutrinos produced by these high-energy primaries. 
    The relevant production channels are shown in Fig.~\ref{fig:decays_Earth_moon}. 
    In our simulation, we take advantage of the fact that the target is a sphere and the cosmic ray flux is isotropic. 
    The setup, shown in Fig.~\ref{fig:sim-geom}, significantly increases the simulation efficiency. 
    The target radius is $r_\mathrm{planet}$, including the atmosphere if it exists.  
    Primary protons are generated in a circular plane with a radius $r_\mathrm{gen}$ that is somewhat larger than $r_\mathrm{planet}$. 
    The protons are generated uniformly along the normal direction (defined as the vertical down direction in Fig.~\ref{fig:sim-geom}). 
    For the Moon, for example, $r_\mathrm{planet} = 1737$ km. 
    We set up detectors at the depth $d$ at all the locations on the sphere with radius $r_\mathrm{planet} - d$ (e.g.~the surface of the yellow ball in Fig.~\ref{fig:sim-geom}). 
    Note for the Earth (or a general target with the atmosphere), $d$ is measured relative to the atmosphere, not  sea level.  
     
    This simulation setup, including the geometry and the radiation sources, is different from the physical situation in which the cosmic ray particles radiate at each location on the surface of the target from every angle isotropically. 
    However, based on the rotational invariance of the system, we can infer the physical flux of a secondary particle at a given depth $d$ and a given location on the full sphere with a radius $r_\mathrm{planet} - d$  from counting all the particles crossing the full surface.  
    We examine energies of neutrinos and muons at all shell crossings. 
    A geometric correction factor is included to reweight those crossings:
    \begin{equation}
        w_g = \frac{\pi\ r_\mathrm{gen}^2}{n_P \ (r_\mathrm{planet} - d)^2\ |\cos(\theta_i)|}
    \end{equation}
    \noindent
    Here, $n_P$ is the total number of simulated primary protons, $\theta_i$ the angle between the secondary particle direction and the radial direction (local zenith angle). 
    The $n_p$ in the denominator ensures the flux results are normalized over the number of primaries generated.
    The $w_g$ factor was checked in the simulation using non-interacting primaries. 
    In this case, the flux at each depth should be the same as the primaries. 

    All nucleons in cosmic rays are treated as protons to a good approximation. 
    For the energy distribution of the primary proton flux. we employ the so-called local interstellar spectra (LIS), i.e., the spectra without solar magnetic field modulation. 
    Therefore, the low-energy secondary particle fluxes could be overestimated in our simulations. 
    Above 100 GeV, it is the same as the flux of galactic cosmic ray flux given by Ref.~\cite{pdg_2022}:
    \begin{equation}
        \Phi_0\ E^{-2.7} \equiv \tilde{\Phi}_{CR}(E) \ . 
        \label{eqn:flux}
    \end{equation}
    Here, $\Phi_0 = 1.7 \times 10^4 /(\mathrm{m}^2\ \mathrm{sr}\ \mathrm{s}\ \mathrm{GeV})$~\cite{Gaisser:2016uoy}, and $E$ is taken in units of GeV. 
    Below these energies, the flux ``rolls off' and flattens. 
    Taking observations from the PAMELA and Voyager 1 spacecraft, Ref.~\cite{bisschoff_2016} gives the following LIS with $E$ under 100 GeV:
    \begin{equation}
        \Phi_p(E) = 3719.0 \frac{1}{\beta^2}\ E^{1.03}\ \bigg( \frac{E^{1.21} + 0.77^{1.21} }{ 1 + 0.77^{1.21}} \bigg)^{-3.18}  \ , \label{eq:lowE_flux_p}
    \end{equation}
    where the cosmic ray intensity $\Phi_p$ has units of [particles m$^{-2}$ s$^{-1}$ sr$^{-1}$ (GeV/nuc)$^{-1}$]; $\beta$ is the  dimensionless particle velocity. The approximate helium LIS is given by
    \begin{equation}
        \Phi_\mathrm{He}(E) = 195.4 \frac{1}{\beta^2}\ E^{1.02}\ \bigg( \frac{E^{1.19} + 0.60^{1.19} }{ 1 + 0.60^{1.19}} \bigg)^{-3.15} \ . \label{eq:lowE_flux_he}
    \end{equation}

    We introduce $w_1$ as the ratio between the LIS flux and the one based on the simple power law $\tilde{\Phi}_{CR}$ in Eq.~\eqref{eqn:flux}:
    \begin{equation}
        w_1 = \frac{4 \Phi_{He} + \Phi_{p}}{\tilde{\Phi}_{CR}} \ . 
    \end{equation}
    Note above 100 GeV, $w_1$ is set to 1, where Eqs.~\eqref{eq:lowE_flux_p} and ~\eqref{eq:lowE_flux_he} lose validity~\cite{bisschoff_2016} but $\tilde{\Phi}$ becomes a good approximation.  
    Therefore, the cosmic ray spectrum from low to high energy can be written as 
    \begin{align}
        \Phi_{CR}(E) = w_1(E) \Phi_0 E^{-2.7} \ . 
    \end{align}

    The simulation generates a set of high-energy proton primaries with $E\in [E_\mathrm{min}, E_\mathrm{max}]$ and $E_\mathrm{min} = 0.3$ GeV, and $E_\mathrm{max} = 30$ TeV. 
    The upper limit was increased as much as our computation time allowed.
    Note that in Fig.~\ref{fig:nuflux_wide}, the maximum energy of the simulated neutrinos is about $10$ TeV.  

    To obtain sufficient simulation statistics across relevant primary energies, the proton primaries were generated following an $E^{-1}$ power law distribution, i.e., 
    \begin{align}
        \rho_{sim}(E) & \equiv C E^{-1} \ ,   \\ 
        C & =   \left(\ln \frac{E_\mathrm{max}}{E_{\mathrm{min}}}\right)^{-1} \ . 
    \end{align}
    Here, $C$ normalizes the distribution to be a probability distribution in the energy range.
    The simulation results are then reweighted by a factor $w_2(E)$ to correct the energy distribution back to that of the primary cosmic ray flux,
    \begin{align}
        w_2(E) & \equiv  \Phi_0  \ln \left(\frac{E_\mathrm{max}}{E_{\mathrm{min}}}\right) E^{-1.7}
    \end{align}
    We ultimately weigh each secondary event by the product $ w_1(E) w_2(E) w_g$ to evaluate the secondary particle fluxes.

\section{Further considerations for semi-analytic estimates} \label{app:semiana}
    
    \begin{table*}[t]
    \centering
    \caption{The key parameters and variables used in analytic flux estimation. 
    Numerical values are adopted from Ref.~\cite{Gaisser:2016uoy}, specifically Tab.~5.2, 5.3 and~5.4 therein. 
    For Tab.~5.2, the ``Sibyll 2.3'' $p$-air values are adopted. $Z_{NN}$ combined both $p$ and $n$ values from Tab.~5.2. 
    $Z_{N\pi}$, $Z_{NK}$ are the sum of the corresponding values for charged mesons in Tab.~5.2. 
    For $\lambda$ and $\Lambda$, their adopted values are those with nucleon energy $E_\text{lab} = 100$ GeV from Tab.5.4. 
    The particle masses are from Ref.~\cite{Workman:2022ynf}.}
    \small
    \begin{tabular}{l c p{8cm} c}
        \hline \hline
        Symbol & Value & Definition/Physical Meaning & Eq. (first occurrence) \\
        \hline
        \multicolumn{4}{l}{\textbf{General}} \\
        $\gamma$ & 1.7 & Integral spectral index of primary cosmic rays & \eqref{eq:primary_flux_scaling} \\
        $\mathcal{C}$ & $1.7$ & Normalization constant ($\rm cm^{-2} sr^{-1} s^{-1} GeV^{-1}$) & \eqref{eq:primary_flux_scaling} \\
        $F_{ji}$ & -- & Energy distribution of the particle $j\to i$ source term  & \eqref{eq:master},\eqref{eq:masterSM} \\
        $x_L$ & -- & Feynman scaling variable ($E_\mathrm{daughter}/E_\mathrm{parent}$) & e.g.,\eqref{eq:ZNN_def} 
        \\
        \hline
        \multicolumn{4}{l}{\textbf{Nucleons (N)}} \\
        $m_N$ & 0.94 GeV & Nucleon mass & -- \\
        $\lambda_N$ & 88 $\units{2}$ & Nucleon interaction length & \eqref{eq:mastereq_N} \\
        $\Lambda_N$ & 120 $\units{2}$ & Nucleon attenuation length: $\lambda_N / (1 - Z_{NN})$ & \eqref{eq:Lambda_Ndef} \\
        $Z_{NN}$ & 0.26 & Spectrum-weighted moment for nucleon regeneration & \eqref{eq:ZNN_def} \\
        \hline
        \multicolumn{4}{l}{\textbf{Pions ($\pi$)}} \\
        $m_\pi$ & 0.14 GeV & Charged pion mass & -- \\
        $c\tau_\pi$ & 780 cm & Pion decay length ($c \times$ lifetime) & -- \\
        $\lambda_\pi$ & 116 $\units{2}$ & Pion interaction length & \eqref{eq:pioneq1} \\
        $\Lambda_\pi$ & 155 $\units{2}$ & Pion attenuation length: $\lambda_\pi / (1 - Z_{\pi\pi})$ & \eqref{eq:pionfluxAtHighE} 
        \\
        $Z_{N\pi}$ & 0.066 & Spectrum-weighted moment for $N \to \pi^\pm$ production & \eqref{eq:Znpidef}
        \\
        $r_\pi$ & 0.573 & Mass ratio squared: $(m_\mu/m_\pi)^2$ & \eqref{eq:dn_dxL_lepton} 
        \\
        $\epsilon_\pi$ & 115 GeV & Critical energy for $\pi$ decay in Earth's atmosphere & \eqref{eq:dpidefgas} \\
        $\tilde{X}_\pi$ & -- & Effective decay length constant in solid ($c \tau_\pi \rho$) & \eqref{eq:dpidefsolid}
        \\
        $Z_{\pi\mu},Z_{\pi\nu}$ & -- & Spectrum-weighted moments for $\pi \to \mu,\nu$ production & \eqref{eq:Z_pimuon_def}\eqref{eq:Z_pinu_def} \\ 
        \hline
        \multicolumn{4}{l}{\textbf{kaons (K)}} \\
        $m_K$ & 0.494 GeV & Charged kaon mass & -- \\
        $c\tau_K$ & 371 cm & kaon decay length ($c \times$ lifetime) & -- \\
        $\lambda_K$ & 134 $\units{2}$ & kaon interaction length & like \eqref{eq:pioneq1} \\
        $\Lambda_K$ & 160 $\units{2}$ & kaon attenuation length & like \eqref{eq:pionfluxAtHighE};\eqref{eq:numuFluxGas}
        \\
        $Z_{NK}$ & 0.011 & Spectrum-weighted moment for $N \to K^\pm$ production & like \eqref{eq:Znpidef}; \eqref{eq:numuFluxGas}
        \\
        $r_K$ & 0.046 & Mass ratio squared: $(m_\mu/m_K)^2$ & \eqref{eq:dn_dxL_lepton} \\
        $\epsilon_K$ & 850 GeV & Critical energy for $K$ decay in Earth's atmosphere & like \eqref{eq:dpidefgas};\eqref{eq:numuFluxGas} \\
        $\tilde{X}_k$ & -- & Effective decay length constant in solid ($c \tau_k \rho$) & like \eqref{eq:dpidefsolid}; \eqref{eq:nuFluxSolid}
        \\
        $Z_{K\mu},Z_{K\nu}$ & -- & Spectrum-weighted moments for $K \to \mu,\nu$ production & like \eqref{eq:Z_pimuon_def}\eqref{eq:Z_pinu_def};\eqref{eq:numuFluxGas} \\ 
        \hline
        \multicolumn{4}{l}{\textbf{Muons ($\mu$)}} \\
        $m_\mu$ & 0.106 GeV & Muon mass & -- \\
        $c\tau_\mu$ & $6.6\times 10^4$ cm & Muon decay length & -- \\
        $\epsilon_\mu$ & 1 GeV & Critical energy for $\mu$ decay in Earth's atmosphere & like \eqref{eq:dpidefgas}; \eqref{eq:Smuon_def} \\
        $\alpha$ & 2.0--2.8 MeV $\units{2}$ & Ionization energy loss & \eqref{eq:muon_E_loss}\eqref{eq:particle_E_loss}\\
        $\xi$ & $2.5 \times 10^5$ $\units{2}$ & Radiative energy loss parameter  &  \eqref{eq:muon_E_loss}\eqref{eq:particle_E_loss} \\
        $\epsilon_\xi$ & $\sim 500$ GeV & Critical energy where radiative losses dominate ($\alpha \xi$) & \eqref{eq:energy_vs_X} \\
        $N_A$ & $6.022 \times 10^{23}/\mathrm{mol}$ & Avogadro's number & \eqref{eq:nu-induced-muon-PDF} \\
        \hline
        \multicolumn{4}{l}{\textbf{Prompt ($\phi$)}} \\
        $\epsilon_\phi,\tilde{X}_\phi,Z_{N\phi},Z_{\phi\nu},Z_{\phi\mu}$ & & similarly defined as the pion counter parts & \eqref{eq:prompt_flux_Solid}-\eqref{eq:Rmu_def} \\
        $R_\nu$ & $\sim 10^{-4}$ & Prompt neutrino flux scaling factor & \eqref{eq:nuFluxSolidPrompt} \\
        $R_\mu$ & $< 10^{-3}$ & Prompt muon flux scaling factor & \eqref{eq:muonFluxSolidPrompt} \\
        \hline
        \multicolumn{4}{l}{\textbf{Environment}} \\
        $X$ & -- & Slant depth (path length $\times$ density) & \eqref{eq:masterSM} \\
        $\theta, X_v$ & -- & Zenith angle, vertical depth ($X_v = X \cos\theta$) & \eqref{eq:PgX}  \\
        $R,g$ & -- & Universal Gas Constant and Gravitational Constant & \eqref{eq:dpidefgas} \\
        $h_0$ & $\sim 6.5$ km & Atmospheric scale height ($RT/Mg$) & \eqref{eq:dpidefgas} \\
        \hline \hline
    \end{tabular}
    \label{tab:params}
    \end{table*}

    In this section, we present our semi-analytic approach for modeling nucleon and charged pion and kaon fluxes in a medium, the meson decays into leptons ($\nu_\mu$, $\mu$), and the prompt component in the lepton fluxes. 
    Then, we discuss $\mu$ energy loss and $\nu_\mu$-induced $\mu$ creation when the leptons travel in the medium.
    The $\mu$ energy loss is crucial for computing $\mu$ fluxes as observed underground (i.e., in the solid), while the $\nu_\mu$-induced $\mu$ production provides the lowest $\mu$ flux underground that cannot be reduced further by going deeper.
    In the end, we discuss the neutrino and muon fluxes in the Moon and Mars. 
    We emphasize that the approach only applies to energetic particles, e.g.~$\nu_\mu$ and $\mu$ with energies above GeV. 
  
    The discussions for fluxes in the gas media mostly follow Chapters~5, 6, and 8 from Ref.~\cite{Gaisser:2016uoy}, which are then generalized for solids. 
    To assist the reader and address the complexity of the parameters, we consolidate all physical parameters, their symbols, and their physical meanings in Table~\ref{tab:params}.
  
\subsection{Charged pions and kaons}\label{app:mesons}

    The master equation system (also see Eq.~\eqref{eq:master}) for computing particle fluxes $N_i(E_i, X)$ can be expressed as 
    \begin{align}
      \frac{dN_i(E_i, X)}{d X} = & -\frac{N_i(E_i,X)}{\lambda_i}-\frac{N_i(E_i,X)}{d_i} \notag \\ 
      & + \sum_{j } \int_E^\infty \, dE_j \frac{F_{ji}(E_i, E_j)}{E_i} \frac{N_j(E_j,X)}{\lambda_j} \,, \label{eq:masterSM}  
    \end{align}
    where $E$ is energy, $X$ is the particle's track length multiplied by the medium mass density (see Table~\ref{tab:geometries} for the values), $i,j$ label particle type, $\lambda_i$ is particle-$i$'s absorption length, and $d_i$ is its decay length. 
    In the right side of the equation, the first two (sink) terms describes the absorption and decay of particle $i$.  
    The last (source) term is for the creation of particle $i$ due to particle $j$'s interaction with medium, with   
    the $F_{ji}(E_i,E_j)$ factor, which is proportional to the corresponding $j\to i$ reaction cross section, characterizing the energy distribution of the source. 
    
    The flux $N_i$ is a double-differential flux, so that 
    $N_i(E_i) d E_i d\Omega_i$ gives the flux within $dE_i$ of energy $E_i$ and within solid angle $d \Omega_i$  of direction $\hat{\Omega}_i$.
    The energy loss of particles are not included here because in gas it is small, while in solid targets the hadronic interaction length is the smallest length scale among all the scales including decay length (for energetic particles) and stopping length (due to energy loss).
    For high-energy particles, the $F_{ji}(E_i, E_j)$ function of two variables can be simplified as a function of $x_L \equiv \frac{E_i}{E_j}$ alone~\cite{Gaisser:2016uoy}. 

    To solve the equations, we need to assign an initial condition for particle fluxes at $X=0$. Here, only the nucleon flux (including the nucleons in cosmic ray nuclei) is nonzero:  
    \begin{equation}
        N_N(E, X=0) = N_0(E) \equiv \mathcal{C}\, E^{-(\gamma+1)} \  ,    \label{eq:primary_flux_scaling}
    \end{equation}
    which is the primary cosmic ray nucleon flux at nucleon kinetic energy $E$. The values for $\gamma$ and $\mathcal{C}$ can be found in Table~\ref{tab:params}.  
    In our estimation, we use $\gamma = 1.7$ for all the energies investigated to simplify the calculations, although for $E$ above $1000\,\mathrm{TeV}$ $\gamma = 2$ is a better approximation.  

    We can solve Eq.~\eqref{eq:master} for particle fluxes when primary nucleons travel through a medium.  
    For nucleons, the equation is decoupled from other hadrons~\cite{Gaisser:2016uoy}:
    \begin{align}
        \frac{d N_N(E_N, X)}{d X} = & - \frac{N_N(E_N, X)}{\lambda_N} \notag \\ 
        & + \int_{E_N}^\infty  d E_N' \frac{F_{NN}(E_N, E_N')}{E_N} \frac{N_N(E_N', X)}{\lambda_N}   \,. \label{eq:mastereq_N}
    \end{align}
    The nucleon flux is then
    \begin{align}
        N_N(E_N,X) =\, & N_0(E_N) e^{- X/\Lambda_N} \ , \text{with}  \notag \\ 
        \Lambda_N \equiv & \frac{\lambda_N}{1-Z_{NN}(\gamma)}     \ ,   \label{eq:Lambda_Ndef} \\ 
        Z_{NN}(\gamma) \equiv & \int_0^1 d x_L\, x_L^{\gamma-1}\, F_{NN}(x_L)   \ . \label{eq:ZNN_def}
    \end{align}
    
    For the pion flux, we have 
    \begin{align}
        \frac{d N_\pi(E_\pi, X)}{d X} = & - \left(\frac{1}{\lambda_\pi} + \frac{1}{d_\pi}\right) N_\pi(E_\pi, X) \notag \\ 
            & + \int_0^1 d x_L \frac{N_\pi(E/x_L) F_{\pi\pi}(E_\pi, E_\pi/x_L)}{x_L^2 \lambda_\pi} \notag \\ 
            & + \int_0^1 d x_L \frac{N_N(E/x_L) F_{N\pi}(E_\pi, E_\pi/x_L)}{x_L^2 \lambda_N} \ . \label{eq:pioneq1}
    \end{align}
    So far, the discussion has applied to both gaseous and solid media, but we will discuss these separately in the following sections because the pion's behavior is quite different in each case. 

    Several length and energy scales are involved in the calculations, including the pion-stopping length (without absorption), decay length, and interaction length. 
    These are depicted in Fig.~\ref{fig:pion_escales}.
    There is a dependence on pion energy in gaseous and solid media, with representative densities at $10^{-3}\,\units{3}$ and $2.8\,\units{3}$, respectively.
    For  gaseous media, pion propagation is dominated by decay and interaction, and thus there is no need to consider pion energy loss. 
    In solids, the dynamics are dictated by the interaction length.
 
    To compute the stopping lengths for charged particles like pions and muons, we solve Eq.~\eqref{eq:muon_E_loss} (or~\eqref{eq:particle_E_loss}) and arrive at Eq.~\eqref{eq:stoppinglength}. 
    Then, based on the fact that $c\,\tau_\pi \approx 8$~m and $\lambda_\pi = 116 \units{2}$ (see Table~\ref{tab:params}), we calculate the decay and interaction lengths respectively. 

    We now examine Eq.~\eqref{eq:pioneq1} for gaseous media, referencing Fig.~5.1 in Ref.~\cite{Gaisser:2016uoy} for the geometrical model of the Earth's atmosphere and the notation used in Eq.~(5.61) therein.
    A key formula is that the vertical length $X_v$ is directly related to the pressure, i.e., 
    \begin{align}
        P = g X_v,  \label{eq:PgX} 
    \end{align}
    with $g$ as the gravitational constant.  
    Meanwhile, the dependence of $X_v$ (and also $P$) on height is an exponential decay to zero when the height approaches infinity. 
    On the Earth at sea level, the vertical depth is $X_0  = 1030\, \units{3}$ (c.f.~Table~A.1 from Ref.~\cite{Gaisser:2016uoy}). 
    Going back to \eqref{eq:pioneq1}, according to Sec.~5.7 in \cite{Gaisser:2016uoy}, we have
    \begin{align}
        \frac{1}{d_\pi} = & \frac{m_\pi\frac{RT}{Mg c \tau_\pi} }{E X \cos\theta} \equiv \frac{\epsilon_\pi}{EX \cos\theta}  \ , \label{eq:dpidefgas}
    \end{align}
    with $M$ as the effective mass of gas molecules averaged across different species, $m_\pi$ the pion mass,  $R$ the universal gas constant, $T$ the atmosphere temperature, $\theta$ the zenith angle of a particle ray for a detector at the surface, and $\epsilon_\pi$ the pion's so-called critical energy. 
    In the same way, $d_K$ and $\epsilon_K$ can be assigned to the charged kaons. 
    The values of $\epsilon_\pi$ and $\epsilon_K$ are provided in Table~\ref{tab:params}. 

    According to Sec.~5.8 in Ref.~\cite{Gaisser:2016uoy}, the solutions to Eq.~\eqref{eq:pioneq1} at both high and low energies take simple forms. When $E \gg \epsilon_\pi$, pion decay can be ignored, and we have 
    \begin{align}
        N_\pi(E,X) = & \,  N(E,0) \frac{Z_{N\pi}}{1-Z_{NN}} \frac{\Lambda_\pi}{\Lambda_\pi - \Lambda_N} \notag \\ 
            & \hspace{2cm} \times \left(e^{-X/\Lambda_\pi}-e^{-X/\Lambda_N}\right) \ , \notag  \\ 
            \text{with}\    \Lambda_\pi = &\, \frac{\lambda_\pi}{1 - Z_{\pi\pi}(\gamma)}  \  \text{and}  \notag \\ 
            Z_{\pi\pi}(\gamma) \equiv & \, \int_0^1 d x_L\, x_L^{\gamma-1} \, F_{\pi\pi}(x_L) \ . \label{eq:pionfluxAtHighE} 
    \end{align}
    The $\lambda_\pi$ and $\Lambda_\pi $ values can be found in Table~\ref{tab:params}.  For $E \ll \epsilon_\pi$, 
    \begin{align}
        N_\pi(E,X) = N(E, 0) \frac{Z_{N\pi}}{\lambda_N} e^{-X/\Lambda_N} \frac{X E\cos\theta}{\epsilon_\pi} \ .  
        \label{eq:pionfluxAtLowE}
    \end{align}
    Here, in the same way as $Z_{NN}(\gamma)$ is defined in Eq.~\eqref{eq:ZNN_def}, the spectrum-weighted moment for the transition $N\to\pi$  is 
    \begin{equation}
        Z_{N\pi}(\gamma) \equiv \int_0^1 d x_L\, x_L^{\gamma-1} \, F_{N\pi}(x_L)   \ . \label{eq:Znpidef}
    \end{equation}   
    The values of $Z_{NN}$ and $Z_{N\pi}$ in Table~\ref{tab:params} are adapted from those of Sibyll 2.3 in the $p$-air column of Table~5.2 of Ref.~\cite{Gaisser:2016uoy}, which takes into account both charged pions. 
    For the estimations in solid media, we continue using the same values for $Z_{NN}$ and $Z_{N\pi}$, since the averaged proton fractions are close to 0.5 in both gaseous and solid media. 

    When computing the pion flux in solids, we do not need to consider energy loss and decay in the medium, as the interaction lengths are far shorter than the other length scales. 
    Therefore, we can apply Eq.~\eqref{eq:pionfluxAtHighE} for energetic pions and similarly for the charged kaons.
    However, we need to examine the charged pion and kaon decays in solid media to compute the lepton fluxes. 
    In this case, the factor $\frac{1}{d_\pi}$ in Eq.~\eqref{eq:pioneq1} is given by 
    \begin{align}
          \frac{1}{d_\pi} = \frac{m_\pi}{E\,  c\,\tau_\pi\, \rho } \equiv \frac{m_\pi}{E \tilde{X}_\pi} 
          \label{eq:dpidefsolid}
    \end{align}
    with $\tilde{X}_\pi$ as the $\pi$'s decay length in the solid. 
    This is similar to \eqref{eq:dpidefgas} with $\epsilon_\pi \to m_\pi$ and $X \to \tilde{X}_\pi$. 
    The $\theta$ dependence is gone, assuming the solid's density to be uniform in all the sites of interest. 
    This term is key for computing lepton fluxes, as discussed next in Sec.~\ref{app:analy_estimate_leptons}. 
    For example, we can see the presence of $\tilde{X}_\pi$ and $\tilde{X}_K$ in the lepton fluxes in Eqs.~\eqref{eq:nuFluxSolid} and~\eqref{eq:muonFluxSolid}.

\subsection{Neutrinos and muons from meson decays} \label{app:analy_estimate_leptons}

    In this section we compute the leptons created from meson two-body decay in both gas and solid. 
    Production from $\mu$ decay and $K_L$ and charged kaons' three-body leptonic decays is not considered, although they are important for estimating the $\nue$ flux in gas. 
    See Fig.~6.7 in Ref.~\cite{Gaisser:2016uoy} for a discussion of these neglected components. 
    Our calculation of $\nu_\mu$ flux in gas is thus good for order-of-magnitude estimates for neutrinos with energies above 1 GeV.  
    Above 100 GeV, it is quite accurate because of the negligibility of $\mu$ decay, see Fig.~6.7 in \cite{Gaisser:2016uoy}.  
  
    For the $\mu$ flux Earth's atmosphere, we must consider muon decay. 
    We know that the hadronic activity peaks about $6.4$ km above sea level. 
    In contrast, the decay lengths of muons with energies around or below 10 GeV are smaller or comparable to the traveled distance if observed at sea level (note that $c\tau_\mu = 0.66$~km). 
  
    In solids, we only consider $\nu_\mu$ neutrinos with energies greater than 1~GeV and $\mu$ with energies at the creation site that are even higher, as we mainly focus on the underground flux. 
    Since the detectors in our study are a few hundred or thousand meters below the surface (much shorter than the decay length), we will ignore the impact of $\mu$ decay on both $\nu_\mu$ and $\mu$ fluxes in solids. 
    Most of the $\nu_\mu$-induced $\mu$ fluxes, relevant deep underground, should come from energetic $\nu_\mu$ because of their larger creation rates.
    The neutrinos from $\mu$ decay are again ignored when computing the $\nu_\mu$-induced $\mu$. 
  
    In short, our estimates of the $\mu$ fluxes from both meson two-body decay and the $\nu_\mu$-induced component should be reliable. 
    The estimation of $\nu_\mu$ fluxes at 1 GeV should be good to within an order of magnitude and becomes more accurate for energies at 10 GeV and beyond. 

    To get the fluxes of daughter leptons from pion and kaon decays, we can compute the production spectrum of leptons at location $X$ as 
    \begin{align}
        \mathcal{P}_i(E,X) = & \sum_{j=\pi,K} \int_{E_{min}}^{E_{max}} d E'\, \frac{F_{ji} (E,E')}{E} \frac{N_j(E', X)}{d_j} \,, \label{eq:prodspectrum}
    \end{align}
    with $i$ denoting lepton species. 
    Properly integrating $\mathcal{P}_i(E,X)$ over $X$ along the meson track will give the lepton flux at a given location. 
    We first examine carefully the integral that, for example, involves $F_{\pi\mu}(E, E')$ for the pion-decay induced $\mu$ flux. Eq.~6.17 in \cite{Gaisser:2016uoy} shows that  
    \begin{align}
        F_{\pi\mu}= F_{\pi\nu} = \frac{1}{1-r_\pi} \ , \text{with} \ r_\pi = \frac{m_\mu^2}{m_\pi^2}\,. \label{eq:dn_dxL_lepton}
    \end{align}
    For kaons, the corresponding $r_K$ is defined in the same way but with $m_\pi \to m_K$.  
    The values of $r_\pi$  and $r_K$ are provided in Table~\ref{tab:params}. 

    Moreover, per Eqs.~\eqref{eq:pionfluxAtHighE} and~\eqref{eq:pionfluxAtLowE}, the pion flux in the integrand scales as $E'^{-(\gamma +1 )}$ and $E'^{-\gamma}$ respectively at high and low energies in the gas, and $E'^{-(\gamma +1 )}$ in the solid. 
    The final piece in the integral, $1/d_\pi$ as a function of $E'$, has a factor of $1/E'$ in both gas and solids based on Eqs.~\eqref{eq:dpidefgas} and~\eqref{eq:dpidefsolid}. 
    After combining all the $E'$ dependence in the integrand, we integrate over the $x_L$ variable defined as $ E/E'$, and arrive at the various moments listed below (see the calculation of $Z_{N\pi}$ in Eq.~\eqref{eq:Znpidef}, now with $N\to \pi$, $\pi \to \mu$, and $N_N/\lambda_N \to N_\pi/d_\pi$ (also see Sec.~6.2 in \cite{Gaisser:2016uoy})). 

    For $\mu$, we define 
    \begin{align}
        Z_{\pi\mu}(\gamma) = \int_{r_\pi}^1 d x_L \frac{1}{1-r_M} x^\gamma = \frac{1-r_\pi^{\gamma+1}}{(1+\gamma)(1-r_\pi)}\,, \label{eq:Z_pimuon_def}
    \end{align}
    for the low energy limit and 
    \begin{align}
        Z_{\pi\mu}(\gamma+1) = \frac{1-r_\pi^{\gamma+2}}{(2+\gamma)(1-r_\pi)} \label{eq:Z_pimuon_def2}
    \end{align}
    for the high energy limit in gas. 
    For $\nu_\mu$, the integration is almost the same except for the integration interval, i.e., $\int_0^{1-r_\pi}$, which leads to 
    \begin{align}
        Z_{\pi\nu}(\gamma) = &  \frac{(1-r_\pi)^\gamma}{(1+\gamma)} \,, \label{eq:Z_pinu_def}  \\ 
        Z_{\pi\nu}(\gamma+1) = &  \frac{(1-r_\pi)^{(\gamma+1)}}{(2+\gamma)} \label{eq:Z_pinu_def2}
    \end{align}
    for the low and high energy limits. 
    The same set of moments, $Z_{K\mu}$ and $Z_{K\nu}$, can be defined for kaon decay.
  
    For solids, we use the moments relevant to the high energy limit (i.e., $Z_{...}(\gamma+1)$) in the flux calculations. 
  
    To get $\nu$ fluxes in gas, we further perform $\int_0^X d X' \, \mathcal{P}_\nu(E, X')$ with $X'=0$ as the initial meson creation location and  $X' = X$ as the location of the detector. 
    Since we already have $\mathcal{P}_\nu(E, X')$ in the high and low energy limits, we can combine them via an interesting interpolation used in Sec.~6.2 in Ref.~\cite{Gaisser:2016uoy}, which gives 
    \begin{align}
        \frac{d N_\nu}{ d E_\nu} = & \frac{N_0(E_\nu)}{1-Z_{NN}} \bigg[ \frac{\mathcal{A}_{\pi\nu}}{1 + \mathcal{B}_{\pi\nu} \cos\theta\, E_\nu/\epsilon_\pi} \notag  \\ 
            & +0.635 \frac{\mathcal{A}_{K\nu}}{1 + \mathcal{B}_{K\nu} \cos\theta\, E_\nu/\epsilon_K} \bigg] \ ,  \  \text{with} \notag \\ 
            \mathcal{A}_{\pi\nu} \equiv & Z_{N\pi} Z_{\pi\nu}(\gamma) \,,\notag  \\ 
            \mathcal{B}_{\pi\nu} \equiv & \frac{Z_{\pi\nu}(\gamma)}{Z_{\pi\nu}(\gamma+1)} \frac{\Lambda_\pi - \Lambda_N}{\Lambda_\pi \ln{\Lambda_\pi/\Lambda_N}} \,,\notag \\ 
            & \text{and similar}\ \mathcal{A}_{K\nu}, \mathcal{B}_{K\nu} \  \text{for charged kaons}\,.\label{eq:numuFluxGas}
    \end{align}
    Note $0.635$ is the charged kaon's branching ratio to the $\mu + \numu$ decay. 
    $N_0(E_\nu)$ is the cosmic ray spectrum. 
  
    For $\mu$ in the atmosphere, the integration over $X'$ must take into account of the $\mu$ energy loss and decay.
    At sea level, we also have similar expressions for the flux:
    \begin{align}
      \frac{d N_\mu}{ d E_\mu} = & S_\mu(E_\mu)\frac{N_0(E_\mu)}{1-Z_{NN}} \bigg[ \frac{\mathcal{A}_{\pi\mu}}{1 + \mathcal{B}_{\pi\mu} \cos\theta\, E_\mu/\epsilon_\pi} \notag \\ 
      & +0.635 \frac{\mathcal{A}_{K\mu}}{1 + \mathcal{B}_{K\mu} \cos\theta\, E_\mu/\epsilon_K} \bigg] \,, \ \text{with} \notag \\ 
      \mathcal{A}_{\pi\mu} = & Z_{N\pi} Z_{\pi\mu}(\gamma) \,,\notag  \\ 
      \mathcal{B}_{\pi\mu} \equiv & \frac{Z_{\pi\mu}(\gamma)}{Z_{\pi\mu}(\gamma+1)} \frac{\Lambda_\pi - \Lambda_N}{\Lambda_\pi \ln{\Lambda_\pi/\Lambda_N}} \,,
      \label{eq:muFluxGas}
    \end{align}
    and similar $\mathcal{A}_{K\mu}, \mathcal{B}_{K\mu}$ for charged kaons.
    The extra factor $S_\mu(E_\mu)$ describes the flux suppression due to muon's energy loss and decay:
    \begin{align}
      S_\mu(E_\mu) = & \left(\frac{\Lambda_N \cos\theta}{X_0}\right)^{p_1} \left(\frac{E_\mu}{E_\mu + \alpha X_0 /\cos\theta}\right)^{p_1 +\gamma + 1 } \notag \\ 
      & \times \Gamma(p_1 + 1) \,,\notag \\ 
      & \text{with}\ p_1 = \frac{\epsilon_\mu}{E_\mu \cos\theta + \alpha X_0}\,.\label{eq:Smuon_def}
    \end{align}
    Here, $X_0$ is the vertical length at sea level, and
    $S_\mu(E_\mu)$ goes to 1 when $E_\mu$ is large enough (a few 10's of GeV, as shown in Fig.~6.2 in \cite{Gaisser:2016uoy}).   
    It should be emphasized that Eq.~\eqref{eq:numuFluxGas} works for $ 0 \leq \theta \leq \pi/2$;
    for $\theta > \pi/2$, we take $\theta \to \pi - \theta$. 
    However, this rule doesn't apply to the muon flux, due to $\mu$ interaction with Earth and its decay. 
  
    For solids, we only use the high-energy limit pion's flux and the $1/d_\pi$ factor in Eq.~\eqref{eq:dpidefsolid}. Note that this $1/d_\pi$ has no $X$ dependence, as opposed to a gaseous medium. 
    As a result, the lepton flux in solids and the corresponding high-energy flux in gases are different, although they share similar pion fluxes as their sources.  
  
    It should be added that although $X$ is integrated from $0$ to $\infty$, physically speaking, the integration is from the contributions within just a few meters below the surface of the solid medium. 
    This is because typical hadron track lengths are on the order of meters in solids, considering the hadronic absorption lengths $\lambda_h\sim 100~\units{3}$.
    After a few meters, all the hadronic activity fades away. 
    In this work,  we consider detectors located a few kilometers or less below the surfaces of solid media. Since the decay lengths of GeV muons are mostly larger than a few kilometers ($c \tau_\mu = 0.66$ km), muon decay can be safely neglected here.  

    We then find for solids, 
    \begin{align}
        \frac{dN_\nu}{d E_\nu} = & \frac{N_0(E_\nu)}{1- Z_{NN}} \Bigg[\frac{\Lambda_\pi m_\pi}{\tilde{X}_\pi E_\nu} Z_{N\pi}Z_{\pi\nu}(\gamma+1) \notag \\ 
        & + 0.635\, \frac{\Lambda_K m_K}{\tilde{X}_K E_\nu} Z_{NK} Z_{K\nu}(\gamma+1) \Bigg] \ , \label{eq:nuFluxSolid}\\
        \frac{dN_\mu}{d E_\mu} = & \frac{N_0(E_\mu)}{1- Z_{NN}} \Bigg[\frac{\Lambda_\pi m_\pi}{\tilde{X}_\pi E_\mu} Z_{N\pi}Z_{\pi\mu}(\gamma+1) \notag \\ 
      & + 0.635\, \frac{\Lambda_K m_K}{\tilde{X}_K E_\mu} Z_{NK} Z_{K\mu}(\gamma+1) \Bigg] \ . \label{eq:muonFluxSolid}
    \end{align}
  
    Interestingly, in the high-energy limit, the ratio between the pion-induced flux in the solid and  the flux in the atmosphere is  
    \begin{align}
        \frac{m_\pi \Lambda_\pi}{\epsilon_\pi \tilde{X}_\pi} = \frac{\Lambda_\pi}{h_0 \rho_\mathrm{solid}} \sim \frac{1 m}{10^4 m} = 10^{-4} \,.\label{eq:gas_solid_flux_ratio_pion}
    \end{align}
    Here, $h_0 = \frac{RT}{Mg}$ is a typical vertical scale of an atmosphere, about $6.5$ km on Earth.  
    Importantly, the flux ratio for the kaon-induced flux has similar sizes, because $\Lambda_\pi \sim \Lambda_K$. 
    Therefore, the gas-to-solid ratio of the combined flux from both mesons is also about $10^{-4}$. 
    However, this ratio is significantly affected by the so-called prompt component that we now discuss.
  
    This prompt component, originating from the decays of charmed mesons and potentially other mesons, dominates the lepton fluxes for energies on the order of 10~TeV and beyond.  
    The lifetimes of these mesons are so short that the decay lengths are the smallest among all the length scales in both gas and solid. 
    Therefore, the flux of these mesons (denoted as $\phi_c$) takes the form of the low-energy pion flux in a gas. 
    Ignoring meson energy loss and its absorption in solids, we get the $\phi_c$ flux,  
    \begin{align}
        \Phi(E,X) = N(E, 0) \frac{Z_{N\phi}}{\lambda_N} e^{- X/\Lambda_n} \frac{\tilde{X}_\phi E}{m_\phi} \,.\label{eq:prompt_flux_Solid}
    \end{align}
    This formula can be derived by substitution in Eq.~\eqref{eq:pionfluxAtLowE}, inspired by the difference between Eq.~\eqref{eq:dpidefsolid} and~\eqref{eq:dpidefgas}. 

    In gas,  
    \begin{align}
        \Phi(E,X) = N(E, 0) \frac{Z_{N\phi}}{\lambda_N} e^{- X/\Lambda_n} \frac{ E X \cos\theta}{\epsilon_\phi} \,.\label{eq:prompt_flux_Gas} 
    \end{align}
    We plug into Eq.~\eqref{eq:prodspectrum} the expressions for $\Phi(E,X)$ and the corresponding $\frac{1}{d_\phi}$, which take the same forms as Eq.~\eqref{eq:dpidefgas} (for gas) and~\eqref{eq:dpidefsolid} (for solid). 
    The lepton fluxes can then be parameterized as 
    \begin{align}
        \frac{dN_\nu^\mathrm{prompt}}{d E_\nu} = & \frac{N_0(E_\nu)}{1- Z_{NN}} \mathcal{A}_{\pi\nu} R_\nu \ , \label{eq:nuFluxSolidPrompt}\\
        \frac{dN_\mu^\mathrm{prompt}}{d E_\nu} = & \frac{N_0(E_\mu)}{1- Z_{NN}} \mathcal{A}_{\pi\mu} R_\mu \ ,  \ \text{with} \label{eq:muonFluxSolidPrompt}  \\ 
        R_\nu \equiv & \frac{Z_{N\phi} Z_{\phi\nu}(\gamma)}{Z_{N\pi}Z_{\pi\nu}(\gamma)} \,,\label{eq:Rnu_def} \\ 
        R_\mu \equiv & \frac{Z_{N\phi} Z_{\phi\mu}(\gamma)}{Z_{N\pi}Z_{\pi\mu}(\gamma)} \,.\label{eq:Rmu_def}
    \end{align}
    These expressions apply to both solids and gases. 
    The ratios $R_\nu$ and $R_\mu$ determine the magnitude of the prompt component. 
    $R_\mu $ is believed to be smaller than $ 2 \times 10^{-3} \equiv \Rmubar$, while $R_\nu $ should be around $ 8 \times 10^{-4} \equiv \Rnubar$ (c.f.~Sec.~8.4 in \cite{Gaisser:2016uoy}). 
    Note that this flux component scales as $E^{-(\gamma +1)}$, which is the same as the low-energy behavior of the flux in the atmosphere, but the pion and kaon-induced fluxes at high energy in the atmosphere and in solids scale as  $E^{-(\gamma +2)}$. 
    Therefore, the prompt component eventually dominates the lepton fluxes when the energy is high enough. 
  
\begin{figure}
        \includegraphics{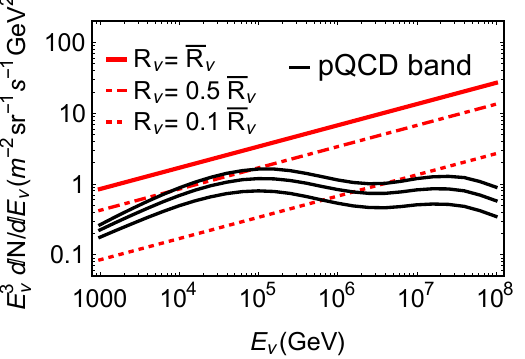}
        \caption{The prompt components in our calculations are compared to the recent pQCD evaluations~\cite{Bhattacharya:2016jce} that have been checked against recent IceCube upper limit~\cite{Aartsen:2016xlq}. }
        \label{fig:Numu_Flux_prompt_component_against_pQCD}
    \end{figure}

    Figure~\ref{fig:Numu_Flux_prompt_component_against_pQCD} examines the effect of the prompt component in the $\nu$ flux on Earth, using three different $R_\nu$ values.
    These were compared against a recent pQCD evaluation~\cite{Bhattacharya:2016jce}, which has been demonstrated to satisfy the IceCube upper limit~\cite{Aartsen:2016xlq}. 
    When $ R_\nu = 0.5 \, \Rnubar$, our parametrization is compatible with the pQCD calculation below  $10^5$ GeV, while $R_\nu = 0.1 \, \Rnubar$ provides reasonable estimates for $E_\nu$ larger than $10^5$ GeV. 
    Fig.~\ref{fig:nuflux_wide} uses these two values to indicate potential variation of our $\nu$ flux predictions in solid.  

    For $\nu_\mu$ fluxes, the formulas presented in the current section are complete, but for $\mu$ fluxes deep underground (with or without atmosphere), we need to consider muon energy loss, which is discussed in the following section.

\subsection{Muon energy loss in solid media}\label{subsec:muonEloss}

    The energy loss of muons through matter can be described by the following formula:
    \begin{align}
        \frac{dE}{dX} = - \alpha - \frac{E}{\xi} \,,\label{eq:particle_E_loss} 
    \end{align}
    with the first term describing the ionization energy loss and the second arising from other processes, including bremsstrahlung, pair production, etc. $\xi$ has mild energy dependence. 
    As an example, for standard rock, $A=22$, $Z=11$, and $\rho = 2.65\, \units{3}$, $\xi = 2.55 \times 10^5\, \units{2}$ at 1~TeV and $2.3\times 10^5\, \units{2}$ at 10~TeV. 
    This energy loss also introduces the so-called critical energy $\epsilon_\xi \equiv \alpha \times \xi$, beyond which the second term in the formula becomes dominant. 
    For muons with $\alpha \sim 2\,\mathrm{MeV}/( \mathrm{g}\,\mathrm{cm}^{-2})$,  $ \epsilon_\xi  = 500$~GeV. 
    
    Based on this formula, we find the particle energy after traveling through length $X$,  
    \begin{align}
        E(X) = (E_0 + \epsilon) e^{-\frac{X}{\xi}} - \epsilon_\xi\,,\label{eq:energy_vs_X} 
    \end{align}
    with $E_0$ as the initial energy. 
    If the muons have energy $E$ as observed at a distance $X$ from the creation point, their initial energy is 
    \begin{align}
        E_0 (E, X) =  e^{\frac{X}{\xi}} \left(E+\epsilon_\xi\right) - \epsilon_\xi \,.
    \end{align}
    For a given $E_0$, the range of the muon, i.e., its stopping length, is then given by  
    \begin{align}
        X^\mathrm{max}(E_0) = \xi \log\left(1+\frac{E_0}{\epsilon_\xi}\right) \ . \label{eq:stoppinglength}
    \end{align}

    The energy spectrum of muons at depth $X$ is completely determined by the flux at $X=0$, as there is a one-to-one correspondence between the energies at different $X$ (the muon's decay can be safely neglected). 
    Therefore, we have 
    \begin{align}
        \frac{d N_\mu (X)}{d E_\mu } = \frac{d N_\mu}{ d E_0} \frac{d E_0}{d E_\mu} = \left. e^\frac{X}{\xi}\frac{d N_\mu}{ d E_0}\right\vert_{E_0 = e^{\frac{X}{\xi}} \left(E+\epsilon_\xi\right) - \epsilon_\xi } \,.
    \end{align}
    This formula predicts the muon flux underground from its flux at sea level on Earth or a few meters below the surface in an environment without atmosphere. 
    The discussion here applies to charged hadrons as well. 
    For example, Eq.~\eqref{eq:stoppinglength} was used to compute the stopping lengths for the charged pions in Fig.~\ref{fig:pion_escales}.

    Note that  $\alpha$ has a mild dependence on muon energy. 
    When $E > 10$ GeV, a good approximation~\cite{Gaisser:2016uoy} is 
    \begin{equation}
        \alpha(E)´\approx  1.9 + 0.08 \ln \left(\frac{E}{m_\mu}\right) \ . 
    \end{equation}
    In our calculations, for simplicity, we apply this dependence for $E > 1$~GeV.  
    In this case $\alpha$ increases slowly to $2.3$ at $10$~GeV, and $2.8$ at $10^4$~GeV. 
    Below $1$ GeV, we set $\alpha = 2$~MeV~$\units{2}$.

\subsection{$\nu_\mu$-induced muons underground}

    To compute this flux, we need to understand the probability of a $\numu$ (with energy $E_\nu$) producing a $\mu$ that has energy between $E_\mu$ and $E_\mu + d E_\mu$ when detected. 
    According to Eq.~8.24 in \cite{Gaisser:2016uoy}, the probability density is
    \begin{align}
        \frac{d \mathcal{P}(E_\nu)}{d E_\mu} = \frac{\xi N_A \sigma_\nu(\epsilon_\xi) }{\epsilon_\xi} \frac{E_\nu-E_\mu}{\epsilon_\xi + E_\mu}\times \begin{cases} 1 \ \text{for $\numu$} \\ 
        \frac{1+ \frac{E_\mu}{E_\nu} +\left(\frac{E_\mu}{E_\nu}\right)^3 }{3} \ \text{for $\numubar$} \end{cases} \  \label{eq:nu-induced-muon-PDF}
    \end{align}
    Here, $\sigma_\nu(\epsilon_\xi)$ is the cross section at energy $\epsilon_\xi$ for $\numu$'s charged-current scattering off nucleons. 
    See Table~\ref{tab:params} for the information about $\epsilon_\xi$, $\xi$ and $N_A$. 
    The derivation depends on the Fermi theory for the weak-interaction cross-section, which holds for $E_\nu$ below 3.6 TeV. 
    In addition, the integrated $X$ between the source of particle production (near the surface) and the detector is required to be on the order of $\xi$ or  larger. 
    If the integrated $X$ is far below this value, for example, just a few hundred meters below the surface in an ice ball, this formula should only be considered as an upper limit. 
    Note the factor $\xi N_A \sigma_\nu(\epsilon_\xi)$  is $ 6 \times 10^{-7}$ when evaluated at $\epsilon_\xi = 500$ GeV.  
  
    The muon flux can be computed via 
    \begin{align}
        \frac{d N_\mu}{d E_\mu} = \int_{E_\mu}^\infty d E_\nu \frac{d N_\nu}{d E_\nu} \frac{d \mathcal{P}(E_\nu)}{d E_\mu} \ . 
    \end{align}
    Note that the probability density for $\numubar$ differs from the $\numu$ one by a factor ranging from $1/3$ to $1$ (see Eq.~\eqref{eq:nu-induced-muon-PDF}). 
    Assuming the same weight for $\numu$ and $\numubar$ in the total neutrino flux, we can multiply the $\nu_\mu$-induced muon flux by a factor between $2/3$ and $1$ to get the full spectrum from neutrinos and antineutrinos. 
    Even without this factor, the neutrino-induced flux is a good estimation of the combined flux from neutrinos and antineutrinos.

\subsection{Fluxes on Moon and Mars}
    Since  atmospheric density is never truly zero, it is natural to ask whether the atmosphere on the Moon or Mars is dilute enough to be considered zero. 
    This question is related to whether meson production and their subsequent decays occur in the atmosphere. 

    The typical length scale of hadronic activity is about 100~$\units{2}$. 
    For example, for high energy pions, the pion flux in \eqref{eq:pionfluxAtHighE} achieves its maximum at 
    \begin{align} 
        X_{max} = \frac{\Lambda_\pi \Lambda_N \ln{\Lambda_N/\Lambda_\pi}}{\Lambda_N  -\Lambda_\pi} \sim 140\,\units{2}
    \end{align}
    Kaons will have similar values. Note when $X\ll X_{max}$, 
    \begin{align}
        \frac{N_\pi(X\ll X_{max})}{N_\pi(X_{max})} = \frac{\left(\frac{1}{\Lambda_N}-\frac{1}{\Lambda_\pi}\right) X}{e^{-\frac{X_{max}}{\Lambda_\pi}} - e^{-\frac{X_{max}}{\Lambda_N}}} \sim \frac{0.02}{\units{2}} X \label{eq:PionfluxratioSmallX}
    \end{align}

    Using Earth's atmosphere as a reference, Eq.~\eqref{eq:PgX} gives the vertical column depths ($X_X$) for the other planets at their surfaces:
    \begin{align}
        X_\oplus = & \frac{P_\oplus}{g_\oplus} \approx 10^4 \, \units{2} \ ,   \\ 
        X_X =  & \frac{P_X/g_X}{P_\oplus/g_\oplus} X_\oplus \notag \\ 
        \approx & \begin{cases} 
        & 1.5 \times 10^{2} \sim X_{max} \  \text{for Mars} \\ 
        & 10^{-10}\sim 10^{-7} \ll X_{max} \  \text{for Moon}
        \end{cases}  \,. 
    \end{align}   
    We then see 
    \begin{align}
        \frac{N_\pi(X_X)}{N_\pi(X_{max})} 
        \approx & \begin{cases}
        &  1 \  \text{for Mars, since}\ X_X\sim X_{max} \\ 
        & 0.02  X_X  \approx 10^{-12}\sim 10^{-9}   
        \end{cases}\,.
    \end{align}
    On the Moon, the pion fluxes on its surface is at most $10^{-9}$ of the maximum flux achieved in Earth's atmosphere.  
    The variation of the ratio for the Moon is caused by the variation of its pressure on the surface. 
    Therefore, the corresponding lepton fluxes are below $10^{-9}$ of the atmospheric fluxes on Earth. 
   
    On the other hand, the pion flux at the surface of Mars is comparable with the maximum flux in Earth's atmosphere, hinting that the lepton fluxes on Mars might be similar to those on Earth. 
    To proceed, we need to understand if pions have enough time to decay in the atmosphere before hitting the surface of Mars. 
    On Earth, $X_{max}$ corresponds to about 14 km above sea level (see Eq.~5.59 in \cite{Gaisser:2016uoy}).
    Pions with energy below 100 GeV are more likely to decay when reaching the surface (note that $c\tau_\pi \approx 8$ m).  
    Since $X_{Mars} \sim X_{max}$, the region where pions achieve maximum flux is around the surface by a physical-length scale defined by $h_0 \equiv RT/Mg$ (note $X_v = X_0 e^{-h/h_0}$ and thus $h_0$ is a typical scale for $X_v$ to vary). 
    On Earth, 
    \begin{align}
        \frac{R T_\oplus}{M_\oplus g_\oplus} \sim 6.5\  \mathrm{km}
    \end{align}
    Since the temperatures on Mars and Earth are on the same scale, $O(10^2)$K, and gas molecule masses are also on the same scale, the ratio between $h_0$ on other planets and  $h_0$ on Earth is proportional to $1/g$. 
    Mars's $g_{Mars}= 0.37 g_\oplus$, meaning $h_0$ on Mars is on the same scale as $h_0$ on Earth. 
    This suggests that pions on Mars have a physical length of at least a km to decay before hitting the surface. 
    Therefore, we find the pion decay physics on Mars is similar to that on Earth, and so are the fluxes of the pion and kaon decay-induced leptons.

\bibliographystyle{apsrev4-1}

\end{document}